\documentclass[aps,twocolumn,superscriptaddress,PRD]{revtex4-2}
\usepackage{amsmath}
\usepackage{epsfig}
\usepackage{multirow}
\usepackage{slashed}
\usepackage{amsmath}
\usepackage{color}
\usepackage[colorlinks,
            linkcolor=blue,
            anchorcolor=blue,
            citecolor=blue
            ]{hyperref}

\def\bea#1\eea{\begin{align}#1\end{align}}

\newcommand{\bef}{\begin{figure}[!htp]}
\newcommand{\eef}{\end{figure}}

\begin{document}
\title{Probing the jet transport coefficient of cold nuclear matter in electron-ion collisions}

\date{\today}

\author{Peng Ru}
\email{p.ru@m.scnu.edu.cn}
\affiliation{School of Materials and New Energy, South China Normal University, Shanwei 516699, China}
\affiliation{Guangdong Provincial Key Laboratory of Nuclear
Science, Institute of Quantum Matter, South China Normal University, Guangzhou 510006, China}
\affiliation{Guangdong-Hong Kong Joint Laboratory of Quantum
Matter, Southern Nuclear Science Computing Center, South China Normal University, Guangzhou 510006, China}

\author{Zhong-Bo Kang}
\email{zkang@ucla.edu}
\affiliation{Department of Physics and Astronomy, University of California, Los Angeles, California 90095, USA}
\affiliation{Mani L. Bhaumik Institute for Theoretical Physics, University of California, Los Angeles, California 90095, USA}
\affiliation{Center for Frontiers in Nuclear Science, Stony Brook University, Stony Brook, New York 11794, USA}

\author{Enke Wang}
\email{wangek@scnu.edu.cn}
\affiliation{Guangdong Provincial Key Laboratory of Nuclear
Science, Institute of Quantum Matter, South China Normal University, Guangzhou 510006, China}
\affiliation{Guangdong-Hong Kong Joint Laboratory of Quantum
Matter, Southern Nuclear Science Computing Center, South China Normal University, Guangzhou 510006, China}

\author{Hongxi Xing}
\email{hxing@m.scnu.edu.cn}
\affiliation{Guangdong Provincial Key Laboratory of Nuclear
Science, Institute of Quantum Matter, South China Normal University, Guangzhou 510006, China}
\affiliation{Guangdong-Hong Kong Joint Laboratory of Quantum
Matter, Southern Nuclear Science Computing Center, South China Normal University, Guangzhou 510006, China}

\author{Ben-Wei Zhang}
\email{bwzhang@mail.ccnu.edu.cn}
\affiliation{Key Laboratory of Quark $\&$ Lepton Physics (MOE) and Institute of Particle Physics, Central China Normal University, Wuhan 430079, China}
\affiliation{Guangdong Provincial Key Laboratory of Nuclear
Science, Institute of Quantum Matter, South China Normal University, Guangzhou 510006, China}

\date{\today}

\begin{abstract}
We present a study of the nuclear-medium induced transverse momentum broadening of particle production in future electron-ion-collision~(EIC) experiments.
By considering the multiple scattering between hard partons and cold nuclear
medium within the higher-twist factorization framework in perturbative QCD, we calculate the
transverse momentum broadening of single hadron production in semi-inclusive measurements, as well as the nuclear enhancement of
the transverse momentum imbalance for di-hadron and heavy-meson pair productions.
In particular, a kinematics dependent non-perturbative jet transport coefficient $\hat q=\hat q(x,Q^2)$
extracted in a global analysis of the current data, together with its uncertainty determined
with a Hessian method, are input into our calculations and are available for the community.
Significant kinematic and color-state dependences of the nuclear induced broadening/imbalance are predicted.
Our results indicate that the future EIC measurements are able to provide powerful constraints
on the kinematic dependence of the transport coefficient $\hat q$ and thus greatly facilitate
the jet tomography of cold nuclear medium.
\end{abstract}

\maketitle

\section{Introduction}
Exploring at a femtometer scale the properties of the nuclear media in different matter phases of quantum chromodynamics~(QCD),
such as cold nucleus, hadron gas and hot/dense quark-gluon plasma, is one of the main goals of various high-energy
nuclear collisions~\cite{Accardi:2012qut,Akiba:2015jwa}, including lepton-, hadron- and nucleus-nucleus collisions.
Benefited by the factorization in perturbative QCD~\cite{Collins:1989gx}, the particle(s) produced with a large momentum transfer,
such as a parton jet, can serve as a well-controlled hard probe of the non-perturbative property of the nuclear medium.
The multiple scattering that a hard probe undergoes in nuclear medium can lead to the transverse momentum broadening and
energy loss of the probe in general~\cite{Gyulassy:2003mc,Cao:2020wlm},
which are reflected in the observed nuclear modifications on the spectra and substructures of jets~(or hadrons)~\cite{Airapetian:2009jy,
Adare:2012qf,Adam:2015jsa,Adams:2003kv,Adler:2005ee,Aad:2010bu,Chatrchyan:2011sx, Adam:2015doa, Chatrchyan:2013kwa,Sirunyan:2018qec,Acharya:2019djg}.
An important medium property commonly embodied in these effects is the jet transport property~\cite{Baier:1996sk,Chen:2011vt, Majumder:2011uk,Burke:2013yra},
quantified as the coefficient $\hat q$,
which characterizes the transverse momentum broadening of a (quark) jet per unit propagation length in the medium and thus measures
the strength of the interaction between the probe and nuclear medium.

Transport coefficient $\hat q$ has been an iconic quantity to represent the medium property seen by jets for a long time,
especially in the study of heavy-ion collisions~\cite{Burke:2013yra, Zhou:2019gqk, Xie:2019oxg, Ma:2018swx, Chen:2016vem, Andres:2016iys,Kumar:2020wvb,JETSCAPE:2021ehl,Xie:2022ght}.
Recently, the study of its dependence on kinematic variables like jet energy and probing scale became active ~\cite{JETSCAPE:2021ehl,Xie:2022ght,Bianchi:2017wpt,Ru:2019qvz,Ru:2020asx,Arratia:2019vju,Kumar:2019uvu,Zhang:2019toi,Arleo:2020rbm,
Bai:2020jmd,Gyulassy:2020jlb,Shi:2019nyp,Zhang:2021tcc},
along with some related theoretical progress~\cite{CasalderreySolana:2007sw,Kang:2013raa,Kang:2014ela,Kang:2016ron,
Blaizot:2014bha,Iancu:2014kga,Liou:2013qya}.
On this aspect, the electron-nucleus~($e$A) and proton-nucleus~($p$A) collisions is of particular importance~\cite{Ru:2019qvz,Ru:2020asx,Arratia:2019vju,Zhang:2019toi,Arleo:2020rbm,Bai:2020jmd,
Kang:2013raa,Kang:2014ela,Kang:2016ron,Kang:2008us,Kang:2012am,Xing:2012ii,Alrashed:2021csd},
since they provide a relative clean environment to delicately study the
kinematic dependence of the transport property of cold nuclear matter and to test the theoretical framework
of the jet-medium interaction, which in turn can be instructive for the study of nucleus-nucleus~(AA) collisions.

In our previous work~\cite{Ru:2019qvz}, we performed the first global extraction of the $\hat q$ in cold nuclear matter
from the current data in $e$A and $p$A collisions, mainly on various types of nuclear-induced transverse momentum broadening.
We found that, with a $\hat q$ depending on the parton momentum fraction $x$ and probing scale $Q^2$,
the theoretical calculations within the higher-twist expansion formalism can give an overall good agreement with the world data.
The extracted optimal $\hat q(x,Q^2)$ shows significant enhancements in small- and large-$x$ regions and a mild $Q^2$ dependence.
However, the uncertainties of $\hat q(x,Q^2)$ were not yet worked out in Ref.~\cite{Ru:2019qvz}. Besides, since most of the
current data are gathered in the intermediate $x$ and $Q^2$ region, the suggested universality and kinematic dependence of
$\hat q(x,Q^2)$ should be examined in future experiments with broader $x$ and $Q^2$ coverage. These issues motivate our followup study presented in this paper.

In the first part of this work, we will upgrade our global analysis of $\hat q$ by determining the uncertainties of the
extracted $\hat q(x,Q^2)$ with the Hessian matrix method~\cite{Pumplin:2000vx,Kovarik:2015cma}, which will generate an uncertainty set of $\hat q(x,Q^2)$.
The Hessian analysis is further upgraded by including a new data set on $J/\psi$ production published recently~\cite{Acharya:2020rvc} to strengthen
the experimental constraints in small $x$ region.
Through the Hessian analysis in this work, we not only learn how the uncertainty of $\hat q(x,Q^2)$ varies with $x$ and $Q^2$,
but also give a complete theoretical prediction with the uncertainty for a related observable, which is important
for the future experimental examination.

The future experiments of electron-ion collisions~(EIC) will be an important place to examine our results~\cite{Accardi:2012qut,Li:2020rqj,Chen:2020ijn}.
Several future EIC facilities have been proposed or under construction, e.g., the
Electro-Ion Collider in US~(US-EIC)~\cite{Accardi:2012qut}, Electron-ion collider in China~(EicC)~\cite{Anderle:2021wcy}
and Jefferson Lab 12~GeV program~(JLab)~\cite{Burkert:2018nvj}, etc. Since these experiments provide a wide
kinematic~($x$ and $Q^2$) coverage and high-precision measurements,
they are expected to significantly improve our understanding of the transport property of cold nuclear matter.

In EIC experiments, the nuclear-induced transverse momentum broadening is still the type of observable most
directly relevant to the transport coefficient $\hat q$. Therefore, in the second part of this work, we study the
transverse momentum broadening of single particle and back-to-back particle pair productions in future electron-ion
collisions. The latter case is also equivalent to the nuclear enhancement of the transverse momentum imbalance of
the particle pair. Our calculations employ the formalism of higher-twist expansion, i.e., the
generalized factorization in perturbative QCD, which has been well established~\cite{Qiu:1990xxa,Qiu:1990xy,Luo:1994np,Luo:1993ui,Luo:1992fz}
and developed~\cite{Kang:2013raa,Kang:2014ela,Kang:2016ron,Kang:2008us,Kang:2012am,Xing:2012ii}, and has shown good
applicability on describing the nuclear modification in both cold and hot/dense nuclear
matter~\cite{Kang:2008us,Kang:2011bp,Burke:2013yra,Ru:2019qvz,Wang:2002ri,Zhang:2003wk}.
In the calculations, we use the $\hat q$ together with its uncertainties extracted from our global analysis as an input.
For comparison, we also provide the predictions with a kinematics independent constant $\hat q$.
By preliminarily estimating the experimental uncertainties, we show the potential of the future EIC measurements to
provide powerful constraints on the $\hat q$ in a wide kinematic range.

The rest of this paper is organized as follows. In Sec.~\ref{sec-qhat}, we briefly review the framework of our global analysis
of $\hat q$ and the main results in our previous analysis~(Sec.~\ref{framework}), and test the extracted $\hat q$ with
a new data set from the ALICE experiment at the Large Hadron Collider (Sec.~\ref{jsi8}).
We then present our work on Hessian analysis of the uncertainty of $\hat q$~(Sec.~\ref{Hessian}).
The jet energy dependence of $\hat q$ is also discussed at the end of this section.
In Sec.~\ref{sec-EIC}, we present the study of three types of nuclear-induced transverse momentum
broadening/imbalance~(Sec.~\ref{sec-sidis}-\ref{sec-ddbar}) for three EIC facilities: US-EIC, EicC, and JLab,
and discuss the advantage of future EICs in understanding the kinematic dependence of $\hat q$~(Sec.~\ref{sec-Reic}).
We give a summary and discussion in Sec.~\ref{sec-summary}.

\section{Global analysis of $\hat q$ for cold nuclear matter}
\label{sec-qhat}
In this section we perform an updated global analysis of the jet transport coefficient $\hat q$ for cold nuclear matter~(CNM) with the current world data from electron-nucleus and proton-nucleus collisions. Through the analysis, a kinematics dependent $\hat q$ is extracted and its uncertainties are estimated with Hessian matrix, which will be used for making predictions for the EIC observables in the next section.

\subsection{Framework and previous results of the analysis}
\label{framework}
First we briefly review the framework of our analysis
as well as what have been done in our previous work~\cite{Ru:2019qvz}.

In high-energy $e$A and $p$A collisions, $\hat q$ is a key
non-perturbative input in theoretical descriptions
of the multiple scattering between the hard probe and the
partons inside the nuclear target.
Generally, in absence of the multiple scattering effects, the cross section of a hard scattering process in $p$A collisions can be written schematically as follows using leading twist collinear factorization formalism
\bea
&~~~~~{d\sigma^S}= f_{q(g)/p}\otimes f_{q(g)/A}\otimes d\hat\sigma^{\textrm{S}}\otimes D_h\,,
\label{eq:fact}
\eea
where the superscript ``$S$" denotes the single-scattering process, $d\hat \sigma^{\textrm{S}}$ represents the perturbatively calculable partonic cross section, $f$ and $D_h$ represent the involved parton distribution functions~(PDFs) in initial state and fragmentation
functions~(FFs) in final state, respectively, and the subscript `${q(g)/A}$' indicates an
incoming quark~(gluon) from the nucleus.
The leading twist collinear factorization formalism underlies the successful global analyses for the PDFs and FFs~\cite{Pumplin:2002vw,deFlorian:2007aj,Kneesch:2007ey, Kovarik:2015cma,Eskola:2016oht,AbdulKhalek:2020yuc}.

In the presence of a large nucleus, the parton multiple scattering effects become important and can be formulated by generalizing the collinear factorization
in the higher-twist expansion approach~\cite{Qiu:1990xxa,Qiu:1990xy,Luo:1994np,Luo:1993ui,Luo:1992fz}.
Specifically, let us consider the transverse
momentum broadening $\Delta \langle p_T^2\rangle$, usually defined as the difference of the averaged transverse momentum square
of the produced particle between $e$A~($p$A) and $ep$~($pp$) collisions,
\bea
\Delta \langle p_T^2\rangle _{eA/pA} = \langle p_T^2\rangle _{eA/pA} - \langle p_T^2\rangle_{ep/pp}\,.
\eea
The leading contribution to the broadening comes from the double scattering effects enhanced by the nuclear size. For example, in semi-inclusive deep inelastic scattering~(SIDIS), the struck quark that is kicked off by the virtual photon may experience additional interactions with the partons inside the nuclear target, resulting in the $\Delta \langle p_T^2\rangle$
of the final-state hadrons.

In higher-twist expansion approach, the leading contribution of transverse momentum broadening can be written generically in the form of a ratio as~\cite{Guo:1998rd}
\bea
\Delta \langle p_T^2\rangle \approx \frac{d\langle p_T^2\sigma^D\rangle}{d\mathcal{PS}}\bigg/\frac{d\sigma^S}{d\mathcal{PS}}\,,
\label{eq:brd1}
\eea
where the denominator ${d\sigma^S}\!/{d\mathcal{PS}}$ is the leading-twist single scattering cross section
in the phase space volume $d\mathcal{PS}$ in $e$A or $p$A collisions, and the numerator
${d\langle p_T^2\sigma^D\rangle}/{d\mathcal{PS}}$ is the $p_T^2$-weighted double scattering cross section~\cite{Kang:2013raa,Kang:2014ela,Kang:2016ron,Kang:2008us,Kang:2012am}
\bea
\frac{d\langle p_T^2\sigma^D\rangle}{d\mathcal{PS}} \equiv \int\!dp_T^2p_T^2\frac{d\sigma^D}{d\mathcal{PS}dp_T^2}\,.
\label{eq:brd2}
\eea
which can be written as follows~\cite{Qiu:2001hj,Qiu:2005ki}
\bea
&{d\langle p_T^2\sigma^D\rangle}=f_{q(g)/p}\otimes T_{ij}\otimes d\hat\sigma^{\textrm{D}}\otimes   D_h\,,
\label{eq:fact4}
\eea
where $T_{ij}$ represents the nuclear twsit-4~(T4) parton-parton correlation functions, which are universal non-perturbative functions encoding the medium properties characterized by the jet transport coefficient $\hat q$.
The $p_T^2$-weighted cross section, thus the broadening, also depends on the color representation
of the hard probe, i.e., the quark and gluon jets correspond to the color factors $C_F=(N_c^2-1)/2N_c$ and $C_A=N_c$,
respectively~\cite{Kang:2013raa,Kang:2014ela,Kang:2016ron,Kang:2008us,Kang:2012am}.

With the assumption of a loosely bound large nucleus, one can neglect the momentum and spatial correlations among the nucleons in the nuclear target. Therefore, the twist-4 matrix element can be effectively factorized in terms of leading twist PDFs and $\hat q$. For example, one can approximate the twist-4 quark-gluon correlation function $T_{qg}$ as~\cite{Kang:2014ela}
\bea
T_{qg}(x,0,0,\mu^2) \approx \frac{9R_A}{8\pi^2\alpha_s} f_{q/A}(x,\mu^2) \hat q(x,\mu^2),
\label{eq:Tqg}
\eea
where $x$ is the momentum fraction carried by the quark that enters into the hard interaction,
$\mu$ is the factorization scale in perturbative QCD, $R_A$ is the radius of the nucleus
with mass number $A$, $f_{q/A}$ is the parton distribution function of quark $q$ in the nucleus, and $\hat q(x,\mu^2)$
is the nuclear geometry averaged jet transport coefficient that we hope to extract.
Similarly, for gluon-gluon correlation function $T_{gg}$, which accesses the
process initiated with a gluon in the nucleus, we assume a same form as Eq.~\eqref{eq:Tqg} in our study,
with $f_{q/A}$ replaced by the gluon distribution $f_{g/A}$.
Through our this paper, the $\hat q$ represents the transport coefficient of a quark jet.

With the theoretical framework introduced above, one can perform a global analysis for $\hat q$
similar to what have been done for PDFs and FFs~\cite{Pumplin:2002vw,deFlorian:2007aj,Kneesch:2007ey, Kovarik:2015cma,Eskola:2016oht,AbdulKhalek:2020yuc}.
In particular, since both the twist-4 correlation functions and the leading-twist PDFs
are universal non-perturbative quantities depend on $x$ and $\mu^2$, the $\hat q$ from Eq.~(\ref{eq:Tqg})
naturally involves possible kinematic dependence on momentum fraction and probing scale.
Note that in phenomenological studies~\cite{Luo:1993ui,Qiu:2003vd,Qiu:2004da} $\hat q$ was usually
assumed to be a constant value due to the unknown kinematic dependence.

Particle transverse momentum broadening is the type of observable most directly relevant to $\hat q$ for CNM~\cite{Qiu:2001hj,Qiu:2005ki}.
Our previous global analysis~\cite{Ru:2019qvz} takes into account the current world data on the transverse momentum broadening of
single hadron production in SIDIS~\cite{Airapetian:2009jy} of $e$A collisions, of Drell-Yan di-lepton production in $p$A
collisions~\cite{McGaughey:1999mq,Bordalo:1987cr}, and of heavy quarkonium~($J/\psi$ and $\Upsilon$) production in
$p$A collisions~\cite{Alde:1991sw,Peng:1999gx,Leitch:1995yc,McGaughey:1999mq,Adare:2012qf,Adam:2015jsa}.
These observables involve multiple scatterings undergone by quark or gluon jets in the initial or~(and) final states
of the hard processes, providing multi-dimensional insight into the transport coefficient as well as a place to
examine the theoretical framework.

Besides the data on transverse momentum broadening,
a set of data on the nuclear modification factor~(shadowing effect) of the DIS structure function~\cite{Adams:1992nf,Adams:1995is}
is also included in our analysis.
In the higher-twist framework, such a nuclear suppression can be related to the coherent multiple scattering, which
has been calculated by resumming the higher-twist contributions~\cite{Qiu:2003vd}, thus, is also sensitive to the transport coefficient.

In total, there are 215 data points in the analysis~\cite{Ru:2019qvz} from the experiments at DESY, FNAL, SPS, RHIC and LHC.
In particular, different observables or measurements involve different kinematic regions identified with
the momentum fraction $x$ of the nuclear parton and the probing scale $Q^2$, providing possibility to explore
the kinematic dependence of $\hat q(x,Q^2)$.

To address the kinematic dependence of $\hat q$ in the global analysis, we adopt the parametrization form~\cite{Ru:2019qvz}
\bea
\hat q(x,\mu^2) = \hat q_0 \,\alpha_s(\mu^2) \,x^{\alpha}(1-x)^{\beta} \left[\ln(\mu^2/\mu_0^2)\right]^{\gamma}\,,
\label{eq-pramt}
\eea
which involves four free parameters, $\hat q_0,~\alpha,~\beta$, and $\gamma$ to be determined by the experimental data.
Such a functional form is primarily motivated with several physical considerations. For example, in the small-$x$ region, we
expect that $\hat q$ depends on the gluon saturation scale, which exhibits a power-law behavior as $Q_s^2 \propto x^{-1/3}$~\cite{GolecBiernat:1998js}. This feature is related to the factor $x^\alpha$ in Eq.~(\ref{eq-pramt}).
At large~$x$, the QCD power corrections could be different~\cite{Dokshitzer:1995qm,Brodsky:2000zu,Gamberg:2014zwa, Braun:2018brg}, and may result in
different behavior of $\hat q$, which is considered by including the factor $(1-x)^\beta$. Moreover, a logarithmic
scale dependence of $\hat q$ is suggested from the radiative corrections~\cite{Iancu:2014kga,Blaizot:2014bha,Collins:2011zzd},
thus we use a factor $[\ln(\mu^2/\mu_0^2)]^{\gamma}$ in Eq.~(\ref{eq-pramt}), where the exponent $\gamma$ is
included to account for potential modification at the higher-order in perturbative corrections and/or non-perturbative contributions.
Although the QCD scale evolution equation of the twist-4 quark-gluon correlation function has been
derived in a series of previous work~\cite{Kang:2013raa,Kang:2014ela,Kang:2016ron}, it is coupled
with the gluon-gluon correlator whose evolution is not determined. In Eq.~(\ref{eq-pramt}), the $\alpha_s(\mu^2)$ is introduced to offset the $\alpha_s$ in the denominator
of the correlation function in Eq.~(\ref{eq:Tqg}), and $\mu_0=1$~GeV is introduced to make the argument of the logarithm dimensionless~\cite{Ru:2019qvz}.
On the whole, the parametrization form in Eq.~(\ref{eq-pramt}) has some generality and is similar to what
is usually used in the extraction of other non-perturbative quantities, such as the parton distribution functions~\cite{Pumplin:2002vw}.
We note that such a parametrization form and several similar forms of $\hat q$ have been recently applied by other groups,
and are shown to work well in their studies~\cite{Arleo:2020rbm,Bai:2020jmd}.

In the analysis~\cite{Ru:2019qvz}, the theoretical calculation of the transverse momentum broadening is performed at
twist-4 level in QCD power expansion and at leading order~(LO) in perturbative expansion with $\alpha_s$.
A complete NLO calculation is not yet available.
In the calculation, we use the CT14 LO parton distribution functions with 3 active quark flavors~\cite{Dulat:2015mca},
and the DSS fragmentation functions~\cite{deFlorian:2007aj}. The heavy quarkonium production is calculated with the
color evaporation model~\cite{Kang:2008us}. Since the transverse momentum broadening is expressed as a ratio in Eq.~(\ref{eq:brd1}),
it is to some extent insensitive to the non-perturbative inputs like the PDFs, FFs, and quarkonium production model.
More details of the calculation can be seen in Ref.~\cite{Ru:2019qvz}.
In addition, the possible hadronization of the in-medium jet may weaken the multiple scattering effects~\cite{Brooks:2020fmf},
which has not been considered in the current framework.

The global analysis of $\hat q$ starts with finding the optimal $\hat q$ by minimizing the $\chi^2$
as a function of the free parameters $\{a_j\}$ defined as~\cite{Pumplin:2000vx,Kovarik:2015cma}
\bea
\chi^2(\{a_j\})=\sum_i\frac{\left[\mathcal{D}_i-\mathcal{T}_i(\{a_j\})\right]^2}{\sigma_i^2},
\label{eq-chi}
\eea
where $\mathcal{D}_i$ is the value of the $i$-th experimental data point, $\mathcal{T}_i(\{a_j\})$ is the corresponding
theoretical prediction depending on the values of the free parameters $\{a_j\}=\{\hat q_0,~\alpha,~\beta,~\gamma\}$ in the
$\hat q$ parametrization in Eq.~(\ref{eq-pramt}), and $\sigma_i^2$ is the statistical and systematic experimental
uncertainties summed in quadrature. The influence of the possible correlated experimental uncertainties has not been taken
into account in our analysis~\cite{Stump:2001gu,Kovarik:2015cma}. The procedure of minimizing the $\chi^2$ in the $\{a_j\}$ space is
performed by utilizing the MINUIT package~\cite{James:1975dr}.

An optimal $\hat q(x,Q^2)$ within the parametrization form in Eq.~(\ref{eq-pramt}) is found at a minimum total $\chi^2=260$~($\chi^2$/NDP$=1.21$,
where NDP$=215$ is the number of data points~\cite{Ru:2019qvz}). The optimal $\hat q(x,Q^2)$ exhibits an obvious $x$ dependence, especially in small- and
large-$x$ regions, as well as a mild $Q^2$ dependence~\cite{Ru:2019qvz}~(also seen in Sec.~\ref{Hessian} of this manuscript).
The theoretical results with this $\hat q(x,Q^2)$ show an overall
good agreement with the experimental data, 
indicating a universal kinematic dependence of $\hat q(x,Q^2)$ in cold nuclear matter.
To further clarify this kinematic dependence, we also performed the fitting by assuming $\hat q$ is a constant quantity
as $\hat q=\hat q_0$, and we found a minimum total $\chi^2=388$~($\chi^2$/NDP$=1.8$), which is apparently larger than that with the kinematic dependence.
Especially, the calculations with the constant $\hat q$ can not give a good description of the data in small- and large-$x$ regions.
For example, the $\chi^2$ for the $J/\psi$ broadening at the LHC is 87.3~($\chi^2$/NDP$=7.3$ with NDP$=12$), which is far from reasonable.
In contrast, the result for this part is significantly improved by using the kinematics dependent $\hat q$,
reflected by the $\chi^2=4.8$~($\chi^2$/NDP$=0.4$)~\cite{Ru:2019qvz}.

The results of the analysis should be examined in future, when more experimental data with a wider kinematic coverage and
a high precision, e.g., from EIC, are available. To this end, one needs the uncertainty of the $\hat q(x,Q^2)$ under
the constraints of the current data, which allows to make a complete theoretical prediction.
In the previous work~\cite{Ru:2019qvz}, a Lagrange multiplier method~\cite{Stump:2001gu,Pumplin:2000vx} is employed
to evaluate the uncertainties of part of the calculated observable.
However, the uncertainties of $\hat q(x,Q^2)$~(varying with $x$ and $Q^2$) can not be easily obtained with that method, which makes the analysis less predictive and thus motivates our
reanalysis with the Hessian matrix method~(Sec.~\ref{Hessian}).
Before presenting that, we show a test for the extracted $\hat q$ with a new data set in the following subsection.

\subsection{A test for $\hat q(x,Q^2)$ with new $J/\psi$ data}
\label{jsi8}
\begin{figure}[b]
\includegraphics[width=3.0in]{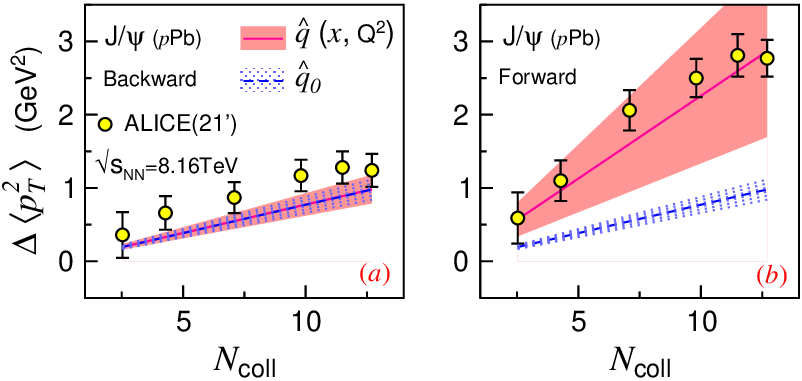}
\caption{Transverse momentum broadening $\Delta \langle p_T^2\rangle$ in $J/\psi$ production in
backward~[panel~(a)] and forward~[panel~(b)] regions in $p$A collisions at the LHC as functions of centrality.
Solid curve represents results with $\hat q(x, Q^2)$ form global analysis~\cite{Ru:2019qvz} and shaded area corresponds to uncertainty.
Dashed line and dotted boundaries represent results with constant $\hat q$ with uncertainty.
Circles are new ALICE data~\cite{Acharya:2020rvc} that are not included in previous analysis of $\hat q$~\cite{Ru:2019qvz}.}
\label{fig:jsi8}
\end{figure}

We noted that the ALICE collaboration have published in 2021 a new data set on the
transverse momentum broadening of $J/\psi$ production in $p$-Pb collisions at $\sqrt{s_{NN}}=8.16$~TeV at the LHC~\cite{Acharya:2020rvc},
which provides an opportunity to test the extracted $\hat q$ in~\cite{Ru:2019qvz}.

In Fig.~\ref{fig:jsi8}, we show both the theoretical results with
the kinematic dependent $\hat q(x,Q^2)$ and the constant $\hat q$ extracted from previous analysis, confronting with the new data.
We find the calculations with the $\hat q(x,Q^2)$ give a visible rapidity dependence of the broadening, reasonably
description of the data in both backward and forward rapidity regions. However, the results with the constant $\hat q$ obviously underestimate the broadening in forward~(small $x$) region.

A similar rapidity dependence can also be observed in the earlier ALICE $J/\psi$ data measured
at $\sqrt{s_{NN}}=5.02$~TeV~(2015)~\cite{Adam:2015jsa}, which have been taken into account in our previous analysis and
provided important information on the $x$ dependence of $\hat q$. In this work, we will also include
the new data in the Hessian analysis to provide more constraints on $\hat q$.

\subsection{Uncertainty of $\hat q(x,Q^2)$ from Hessian Matrix}
\label{Hessian}
In order to estimate the uncertainties of the kinematics dependent $\hat q(x,Q^2)$ under the constraints of the current data,
we perform an analysis with a Hessian matrix method. The basic assumption of the Hessian analysis is that the $\chi^2$ can be
approximately expressed in a quadratic form of the free parameters $\{a_i\}$ in the neighborhood of the minimum~\cite{Pumplin:2000vx,Kovarik:2015cma} as
\bea
\chi^2=\chi_0^2+\sum_{i,j}H_{ij}y_iy_j\,,
\eea
where $\chi_0^2\equiv\chi^2(\{a_i^0\})$ is the global minimum of the $\chi^2$ at the optimal parameter values $\{a_i\}=\{a_i^0\}$,
$y_i=a_i-a_i^0$ is the displacement of $a_i$ from its optimal value $a_i^0$, and $H_{ij}$ are the elements of Hessian matrix defined as
\bea
H_{ij}=\frac{1}{2}\left(\frac{\partial^2\chi^2}{\partial y_i\partial y_j}\right)_{a_i=a_i^0}.
\eea
Usually there are interplays among different variables $y_i$~(or $a_i$) in the $\chi^2$, and the off-diagonal Hessian
matrix elements could be non-zero. This makes the uncertainty estimation, corresponding to a certain tolerance
$\Delta\chi^2\equiv\chi^2-\chi^2_0$, not that straightforward.
However, one can disentangle the parameters by defining a new basis $\{z_i\}$ of the parameter space,
in whose representation the Hessian matrix is diagonal. With the new set of parameters $\{z_i\}$,
the $\Delta\chi^2$ can be written in a simple form as
\bea
\Delta\chi^2=\sum_i z_i^2,
\eea
which means that the contours of the $\chi^2$ are spheres in the new basis. Using this new basis, one can
generate the corresponding uncertainty sets of $\hat q$,
which can be used to estimate both the uncertainties of $\hat q$ and the related theoretical predictions.
More details of the Hessian analysis can be found in Appendix~\ref{section:appendix}.

\begin{figure}[t]
\includegraphics[width=2.6in]{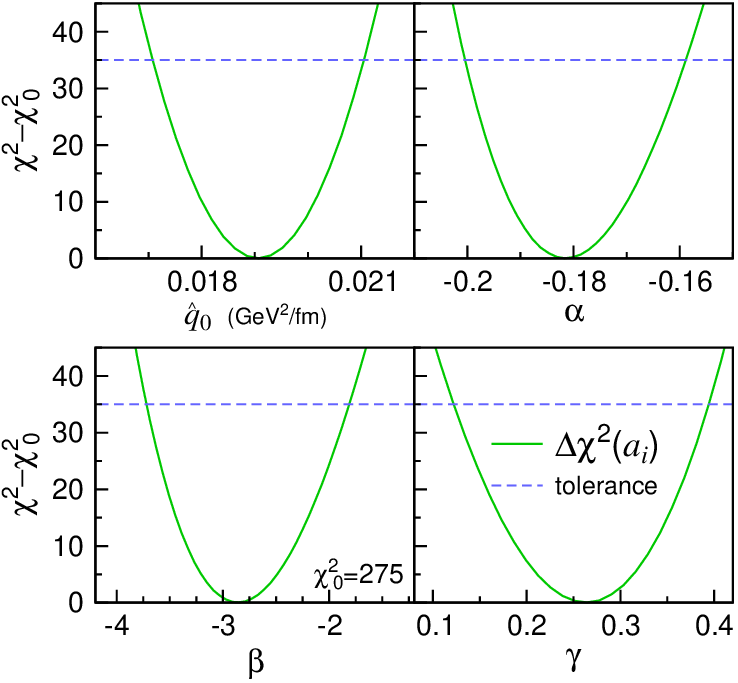}
\caption{$\Delta\chi^2=\chi^2-\chi^2_0$ as a function of each individual original parameter $a_i$ in neighborhood
of minimum $\chi^2=\chi^2_0$. Here $\{a_i\}=\{\hat q_0, \alpha, \beta, \gamma\}$. Dashed line represents
tolerance $\Delta\chi^2=35$.}
\label{fig:chi2a}
\end{figure}

The Hessian analysis in this work is performed by using the MINUIT package combined with the ITERATE program~\cite{Pumplin:2000vx}.
The theoretical framework for calculating various observables is the same as in our previous analysis introduced in Sec~\ref{framework}.
After the new ALICE data on $J/\psi$ production~(Sec.~\ref{jsi8})
included~(now 227 data points in total), a global minimum $\chi_0^2=275$ is reached through the analysis.

We first show in Fig.~\ref{fig:chi2a} the global $\chi^2-\chi_0^2$ as functions of each individual original parameters $\{a_i\}$
in the vicinity of the minimum. We can see that the experimental data indeed have sensitivities to all the parameters.
The non-zero optimal values of the parameters $\alpha$, $\beta$ and $\gamma$ suggest the kinematic dependence of $\hat q$
on $x$ and $Q^2$, to be further consolidated with the determined uncertainty.

\begin{figure}[t]
\includegraphics[width=2.6in]{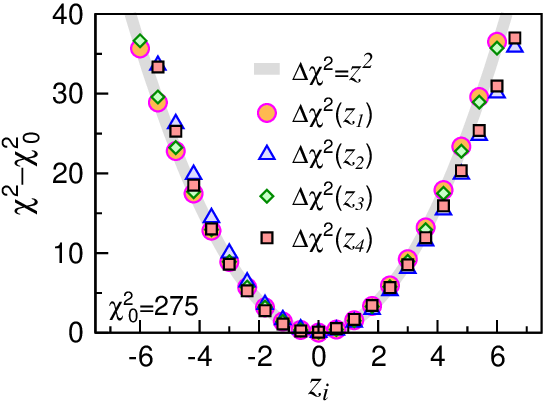}
\caption{$\Delta\chi^2=\chi^2-\chi^2_0$ as functions of new parameters $\{z_i\}$ from Hessian analysis.
Grey curve shows a quadratic function $\Delta\chi^2=z^2$ for reference.}
\label{fig:chi2z}
\end{figure}

\begin{figure*}[t]
  \includegraphics[scale=0.6]{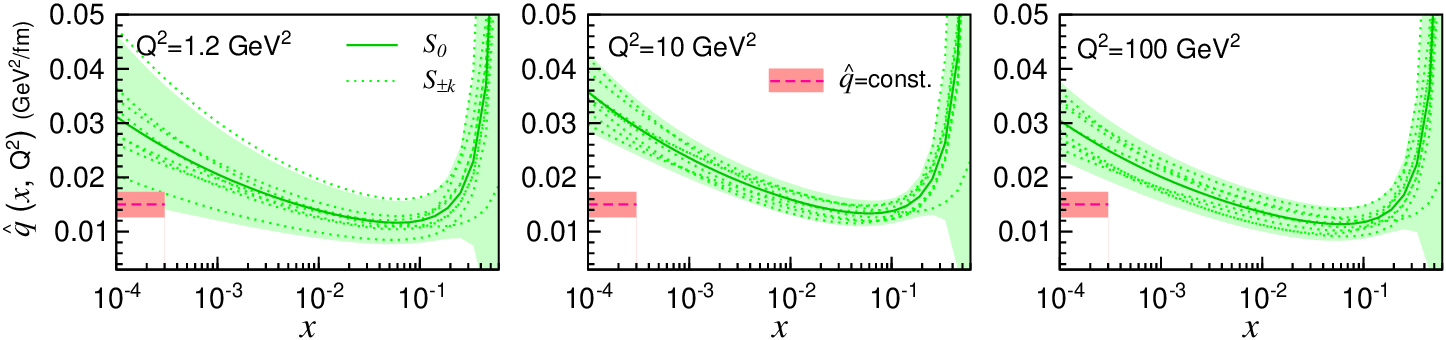}\\
  \caption{$\hat q(x,Q^2)$ extracted from global analysis, shown as a function of momentum fraction $x$ of initial-state nuclear
  parton at $Q^2=$1.2, 10, and 100~GeV$^2$. Green solid curve represents optimal values~($S_0$), light-green band represents corresponding
  uncertainty, and dotted curves show uncertainty set $\{S_{\pm k}\}$ of $\hat q(x,Q^2)$. For reference, kinematics independent constant $\hat q$
  and uncertainty extracted from data are shown as red dashed line and pink band, respectively.
  Both uncertainties of $\hat q(x,Q^2)$ and constant $\hat q$ correspond to $90\%$ C.L..
  }\label{fig:qk_unc}
\end{figure*}

\begin{figure}[t]
\includegraphics[width=2.3in]{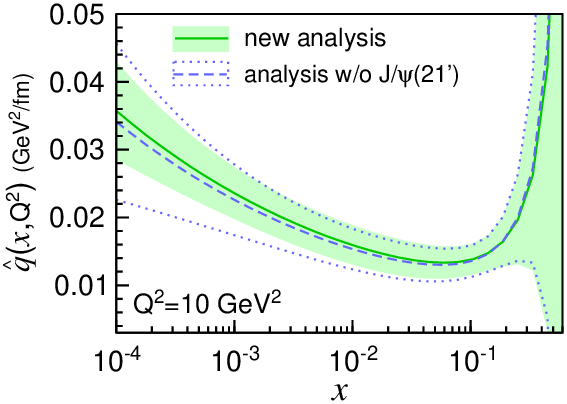}
\caption{$\hat q(x,Q^2)$ at $Q^2=10$~GeV$^2$ and uncertainties extracted from global analysis with and without ALICE data~(2021)
on transverse momentum broadening of $J/\psi$ production. Results of new analysis are the same as in Fig.~\ref{fig:qk_unc}.
Results without $J/\psi$~(21') data are shown with a dashed curve with dotted boundaries.
}
\label{fig:qhat_unc_81}
\end{figure}

Fig.~\ref{fig:chi2z} shows the values of $\chi^2-\chi_0^2$ as functions of the new parameters $\{z_i\}$ defined in Eq.~(\ref{eq-zi}).
They are found to have a very good agreement with the quadratic form $\Delta\chi^2=z_i^2$, indicating the good performance of
Hessian analysis.

With this Hessian analysis, we obtain the optimal values of the parameters $\{a_i\}$ together with their uncertainties
corresponding to $90\%$ confidence level~(C.L.):
\begin{eqnarray}
\hat{q}_0=0.0191\!\pm 0.0061~\textrm{GeV}^2\!/\textrm{fm},~~ \alpha=-0.182\pm 0.050,\cr
\beta=-2.85\pm 1.87,~~~ \gamma=0.264\pm 0.169\,.~~~~~~
\label{eq-para}
\end{eqnarray}
We find that with the uncertainties, the global data favor negative $\alpha$ and $\beta$, and a positive $\gamma$ in the parametrization of
$\hat q$. The optimal values of the parameters are in good agreement with the results of our previous analysis~\cite{Ru:2019qvz}.

\begin{figure}[b]
\includegraphics[width=3.0in]{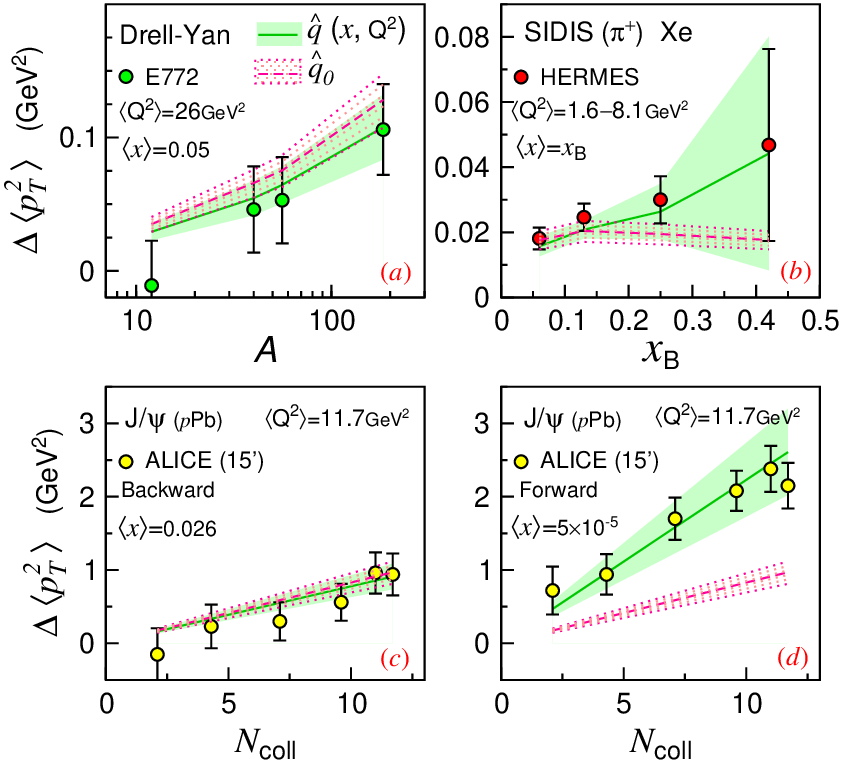}
\caption{Transverse momentum broadening $\Delta \langle p_T^2\rangle$, in Drell-Yan process in $p$A collisions versus
nuclear mass number $A$~[panel~(a)], in SIDIS versus Bjorken $x_B$~[panel~(b)], and in $J/\psi$ production in
backward~[panel~(c)] and forward~[panel~(d)] regions in $p$A collisions at the LHC~(5TeV, 2015) as functions of $N_{\textrm{coll}}$.
Green solid curve represents results with optimal $\hat q(x, Q^2)$ and light-green shaded area corresponds to uncertainty.
Red dashed line and dotted band represent results with constant $\hat q$ with uncertainty.
Averaged momentum fraction $\langle x\rangle$ and probing scale $\langle Q^2\rangle$ are shown for reference.}
\label{fig:qkq0}
\end{figure}

In Fig.~\ref{fig:qk_unc}, we show the extracted $\hat q(x, Q^2)$ versus the momentum fraction $x$ for $Q^2=$1.2, 10, and 100~GeV$^2$,
including the optimal values~($S_0$), uncertainty bands, and uncertainty sets $S_{\pm k}$ of $\hat q(x, Q^2)$~[defined by Eq.~(\ref{eq-qk})],
which can be used to estimate the uncertainty for theoretical predictions~[see Eq.~(\ref{eq-unc})].
For the optimal values, we can see that the $x$ dependence is noticeable in small- and large-$x$ regions, and the $Q^2$ dependence
is relatively mild. The enhancements of $\hat q(x, Q^2)$ in small- and large-$x$ regions are related to the negative parameters
$\alpha$ and $\beta$, respectively. The $\alpha$ value in Eq.~({\ref{eq-para}}) is qualitatively consistent with the
growth rate of the gluon density expected in saturation physics~\cite{Iancu:2003xm},
and the negative $\beta$ may indicate an enhancement of the nuclear power correction
at large $x$~\cite{Dokshitzer:1995qm,Brodsky:2000zu,Gamberg:2014zwa, Braun:2018brg}.
Since most of the current data are located in the intermediate $x$ and $Q^2$ regions, the uncertainties of
$\hat q(x, Q^2)$ become larger at small and large values of $x$ or $Q^2$, due to the less experimental constraints.
Especially, the uncertainties are dramatically large at $x\gtrsim0.4$, where no data exist for this kinematic region.
For comparison, we also show in Fig.~\ref{fig:qk_unc} the $\hat q$ extracted by assuming it is a kinematic independent
constant quantity~($\hat q= \hat q_0=0.0150^{+0.0023}_{-0.0025}$~GeV$^2$/fm).

\begin{figure*}[t]
\includegraphics[width=5.0in]{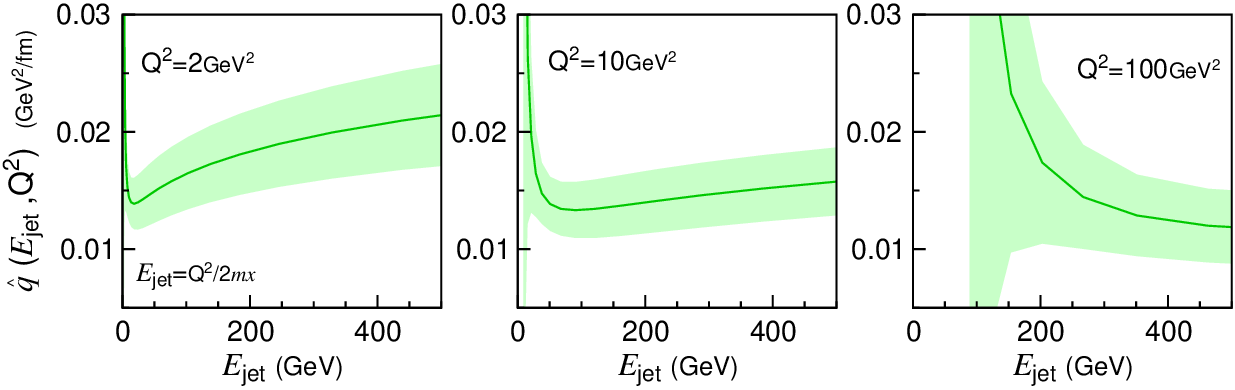}
\caption{Extracted $\hat q$ as a function of jet energy $E_{\textrm{jet}}$.
Solid curve and shaded band represent optimal values and uncertainties, respectively.
}
\label{fig:Ejet}
\end{figure*}

To illustrate the impact of the added new $J/\psi$ data on the extracted $\hat q(x,Q^2)$, we compare in
Fig.~\ref{fig:qhat_unc_81} the $\hat q(x,Q^2)$ for $Q^2=10$~GeV$^2$ extracted with and without the new $J/\psi$ data.
A good agreement between their optimal values is found.
What is notable is that, with the additional constraints from the new data, the uncertainties at small values of $x$ are reduced to
some extent and a more evident $x$ dependence is observed. More detailed illustrations for the impact of the new $J/\psi$ data on $\hat q(x,Q^2)$
at different $Q^2$ and on the predictions for different observables can be seen in Appendix~\ref{section:appendix2}.
In addition, for theoretical predictions, we have compared the uncertainties from the Hessian analysis with those from our previous analysis via
the Lagrange multiplier method, and found a good agreement between them~\cite{Ru:2021hwv}.

\subsection{Kinematic dependence of $\hat q$}

The kinematic dependence of $\hat q$ is of particular interest since it is related to
the detailed partonic structures of the nuclear matter, similar to the $x$ and $Q^2$ dependence
of a parton distribution function. Global analysis offers an indispensable data-driven understanding of such issues.
To demonstrate how the experimental data determine the kinematic dependence of the $\hat q$ in our analysis,
we show in Fig.~\ref{fig:qkq0} the results calculated with the extracted $\hat q(x, Q^2)$ and constant
$\hat q= \hat q_0$, for four representative observables in the analysis: the transverse momentum broadening in Drell-Yan
process~(nuclear mass number dependence), in SIDIS~(Bjorken $x$ dependence), and in $J/\psi$ production~[dependence on
the collision-centrality related $N_{\textrm{coll}}$ in backward
and forward rapidity at the LHC].
For the Drell-Yan~[panel~(a)] and the backward $J/\psi$ production~[panel~(c)], since the involved momentum fraction $x$ of the nuclear parton
is in the intermediate region, the results with the $\hat q(x, Q^2)$ and the constant $\hat q$ are close to each other, while a slightly better
agreement with the Drell-Yan data is given by that with the $\hat q(x, Q^2)$.
However, the theoretical predictions are significantly improved with the kinematics dependent $\hat q(x, Q^2)$ in the SIDIS~[panel~(b)]
and forward $J/\psi$ production~[panel~(d)], which correspond to the regions of large and small $x$, respectively.
In particular, the calculation with the constant $\hat q$ from the global analysis completely fails to describe the data
on the forward $J/\psi$ production at the LHC.
Actually, the enhancements of the $\hat q(x, Q^2)$ in small and large $x$ regions observed in Fig.~\ref{fig:qk_unc}
stem largely from the $J/\psi$ production in forward region at the LHC and the Bjorken-$x$ dependence of SIDIS in the analysis, respectively,
which should be examined through future experiments that involve small and large $x$ regions.

On the other hand, it is also noteworthy that, for the observables that are taken into account in our analysis, i.e.,
the transverse momentum broadening in SIDIS, Drell-Yan process and heavy-quarkonium production,
the energy of the hard probe $E_{\textrm{jet}}$ in the nucleus rest frame can be commonly expressed with the Lorentz invariant variables $x$ and $Q^2$ through the relationship:
\bea
E_{\textrm{jet}}=\frac{Q^2}{2m_px}\,,
\label{eq-Ejet}
\eea
where the hard scale $Q^2$ is the virtuality~(or squared invariant mass) of the virtual photon in SIDIS and Drell-Yan process,
and is the squared invariant mass of the heavy-quark pair that form the quarkonium in the color evaporation model, $x$ is the momentum fraction of the initial-state nuclear parton in these processes, and $m_p$ is the nucleon mass.
With Eq.~(\ref{eq-Ejet}), we can convert the extracted $\hat q(x, Q^2)$ into the form $\hat q(E_{\textrm{jet}}, Q^2)$,
which can be regarded as the jet energy dependence of $\hat q$ in cold nuclear matter. Since the jet energy dependence
is usually discussed in the study of jet quenching in quark-gluon plasma in relativistic heavy-ion
collisions~\cite{Zhou:2019gqk,CasalderreySolana:2007sw,JETSCAPE:2021ehl}, to extract
the $\hat q(E_{\textrm{jet}}, Q^2)$ in cold nuclear matter will provide a reference for the future comparative study.
In Fig.~\ref{fig:Ejet} we plot the $\hat q(E_{\textrm{jet}}, Q^2)$ with uncertainties. We find that for $Q^2=2-10$~GeV$^2$
the $\hat q$ increases with $E_{\textrm{jet}}$ in a wide range of jet energy corresponding to the small $x$ region.
However, for $Q^2=100$~GeV$^2$, the plotted region corresponds to large $x$ values, and the $\hat q$ decreases with
 $E_{\textrm{jet}}$ with a large uncertainty.
Some similar results in the study of jet quenching can be found in Refs.~\cite{Zhou:2019gqk,CasalderreySolana:2007sw,JETSCAPE:2021ehl}.
Besides, the $Q^2$ dependence in $\hat q(E_{\textrm{jet}}, Q^2)$ is more pronounced than that in $\hat q(x, Q^2)$, because the corresponding $x$ will
vary with $Q^2$ for a certain $E_{\textrm{jet}}$.

Trough the global analysis discussed in this section, a kinematics dependent transport coefficient $\hat q$ in cold
nuclear matter and its uncertainties have been extracted from the current experimental data, which is expected to motivate
the future theoretical and experimental studies to further understand and constrain the $\hat q(x, Q^2)$,
and to consolidate the universality of the kinematic dependence of $\hat q$. Next, we will study related observables
in future electron-ion collisions.

\section{Nuclear induced transverse momentum broadening/imbalance in electron-ion collisions}
\label{sec-EIC}
The future EIC facilities~\cite{Accardi:2012qut,Anderle:2021wcy,Burkert:2018nvj} provide great opportunities to deepen our understanding of the jet transport property of the cold nuclear medium.
In EIC experiments, the nuclear-medium induced transverse momentum broadening will continue to be the observable
most directly related to the transport coefficient $\hat q$. Since the initial-state projectile is an electron in EIC,
the broadening is induced by the final-state multiple scattering between the outgoing hard probe and the nucleus.

In this section, we will study the transverse
momentum broadening in both single- and pairwise-particle productions at the EIC. For the latter case, the
nuclear broadening of the particle-pair is equivalent to the nuclear enhancement
of the particle-pair transverse momentum imbalance.
Concretely, using the $\hat q$ extracted in the previous section, we will calculate the transverse
momentum broadening of single-hadron production and the enhancement of transverse momentum imbalance of di-hadron and
heavy-meson pair~($D\bar D$) productions. These observables will be studied in the kinematic regions of
three proposed EIC facilities: US-EIC, EicC, and JLab~(12GeV)~\cite{Accardi:2012qut,Burkert:2018nvj,Anderle:2021wcy}.

Since we focus on the EIC, it will be useful to give the typical Lorentz-invariant DIS kinematic variables
\bea
x_B=\frac{Q^2}{2p_N\cdot\!q_{\gamma}}\,,~~ y=\frac{q_{\gamma}\cdot\!p_N}{k_e\cdot\!p_N}\,,~~
Q^2=-q^2_{\gamma}\,.
\eea
Here $x_B$ is the Bjorken variable, $y$ is the inelasticity of the scattering, and $Q^2$ is the virtuality of the
exchanged photon $\gamma^{\ast}$. In their definitions, $k_e$, $p_N$ and $q_{\gamma}$ are the 4-momenta of the incoming electron,
the nucleon and the virtual photon, respectively. For the final-sate fragmentation process, the hadron momentum fraction $z_h$
is introduced as
\bea
z_h=\frac{p_N\cdot\!p_h}{p_N\cdot\!q_{\gamma}},
\eea
where $p_h$ is the 4-momentum of the final-state hadron. In addition, the squared invariant mass of the
photon-nucleon~($\gamma^{\ast}$-N) system is $W^2=(q_{\gamma}+p_N)^2$.

\begin{figure}[t]
\includegraphics[width=2.3in]{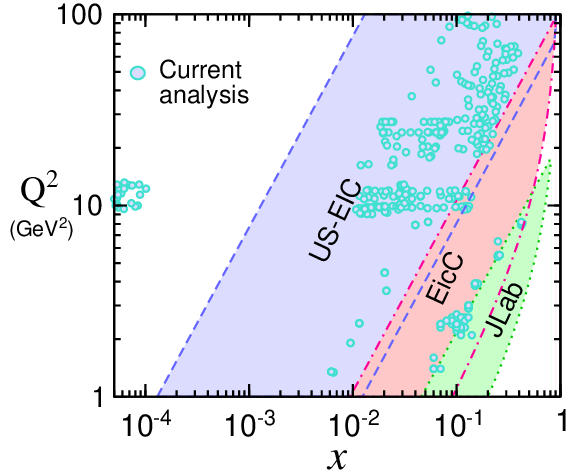}
\caption{Kinematic~($x_B$ and $Q^2$) regions covered by three EIC facilities shown as shaded area.
Boundaries for US-EIC, EicC, and JLab are plotted with dashed, dotted-dashed, and dotted curves, respectively.
Discrete circles are sampled momentum fraction $x$ and probing scale $Q^2$ from numerical calculations in current
analysis of $\hat q$~(Density of the circles doesn't represent the density of data points).}
\label{fig:EIC_kinematic}
\end{figure}

In our calculations, the center-of-mass energies~($\sqrt{s}$) of the electron-nucleon system for the three EIC
facilities are taken to be 90~GeV~(US-EIC), 10.6~GeV~(EicC), and 4.8~GeV~(JLab), respectively~\cite{Accardi:2012qut,Burkert:2018nvj,Anderle:2021wcy, Aschenauer:2017oxs}.
The kinematic range considered in our calculations is: $Q^2>1$~GeV$^2$,
$0.01<y<0.95$, $W^2>10$~GeV$^2$~($>4$~GeV$^2$ for JLab), and $0.3<z_h<0.8$~\cite{Aschenauer:2017oxs,Aschenauer:2019kzf}.
With this kinematic restriction, we plot in Fig.~\ref{fig:EIC_kinematic} the ranges of Bjorken $x$ and $Q^2$
covered by the three facilities. For comparison, some sampled values of the $x$ and $Q^2$ involved in our global
analysis of the current data are also shown. We can see that the future EIC facilities have the
potential to allow a high-coverage scan on the kinematic dependence of $\hat q$, especially for small-
and large-$x$ regions where the current measurements rarely access.

Besides the wide kinematic coverage, the high precision measurements at future EIC are expected to provide
more powerful constraints on the $\hat q$. In this study, to preliminarily estimate the uncertainty of the measurement in
future EIC, we consider an integrated luminosity as $\mathcal{L}=5$~fb$^{-1}$~\cite{Burkert:2018nvj,Anderle:2021wcy,Aschenauer:2017oxs},
and evaluate the relative statistical uncertainty as $\delta_{st}=1/\sqrt{\sigma\mathcal{L}}$, where
$\sigma$ is the cross section of the considered process. Due to the lack of the information on the
systematic uncertainty, we simply assume it is on the same order of the statistical uncertainty, and include an
additional factor $\sqrt{2}$ to estimate the total uncertainty.
Now we present the study for three types of nuclear induced broadening/imbalance as follows.

\subsection{Single hadron $p_T$ broadening in SIDIS}
\label{sec-sidis}
\begin{figure*}[t]
\includegraphics[width=5.0in]{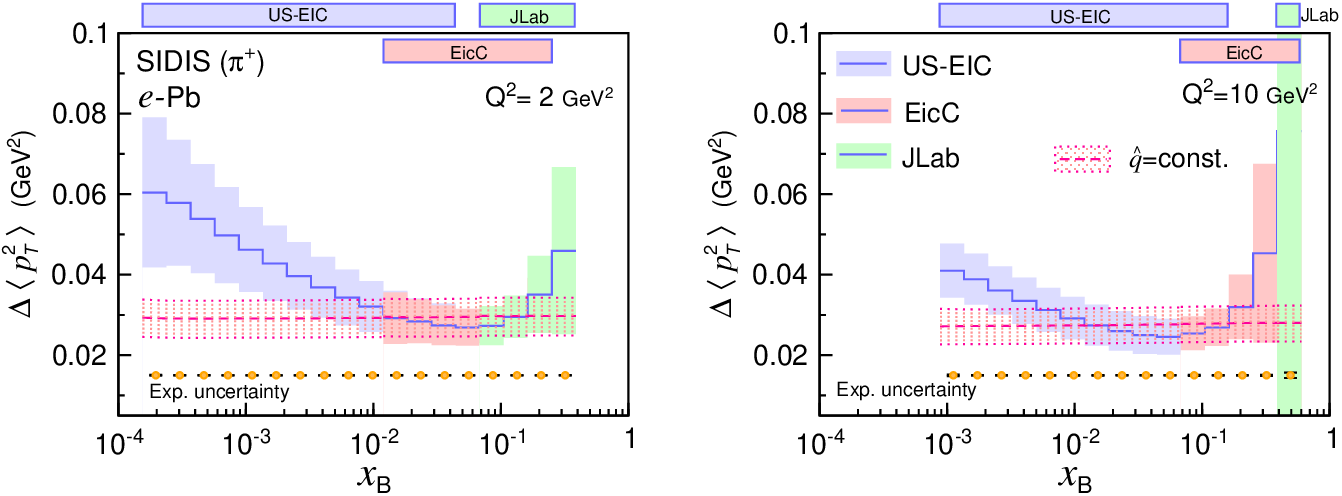}
\caption{Transverse momentum broadening $\Delta \langle p_T^2\rangle$ of single pion production in SIDIS as a function of Bjorken
$x_B$ at $Q^2=$2 and 10~GeV$^2$, for three EIC facilities. Solid curve with shaded area represents results with $\hat q(x,Q^2)$
with uncertainties. Dashed curve with dotted band shows results with constant $\hat q$ with uncertainties.
Circles with vertical bars represent estimated experimental uncertainties.
In calculations, we have taken $Q^2\in[1.5,2.5]$~GeV$^2$~(left panel) and $Q^2\in[9,11]$~GeV$^2$~(right panel), and set $x_B<0.6$.
On top of each panel, we mark the kinematic~($x_B$) ranges covered by three facilities, and overlaps among them can be seen.
The theoretical results for two facilities in their overlap region is generally similar, since they depend on
$x_B$ at $Q^2$ to a large extent.
}
\label{fig:sidis_xb}
\end{figure*}

\begin{figure}[t]
\includegraphics[width=2.3in]{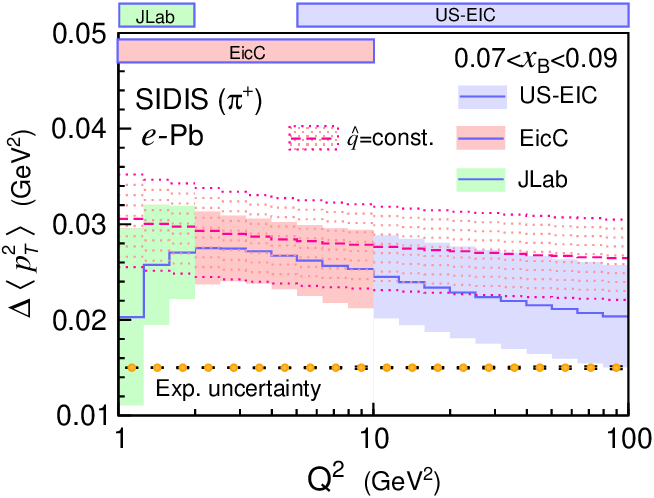}
\caption{Similar as Fig.~\ref{fig:sidis_xb}, but for transverse momentum broadening $\Delta \langle p_T^2\rangle$ of
single pion production in SIDIS as a function of $Q^2$ within $0.07<x_B<0.09$.
Results are calculated for $1<Q^2<100$~GeV$^2$. }
\label{fig:sidis_q2}
\end{figure}

The transverse momentum broadening of the single hadron production in SIDIS is a type of observable that has already
played an important role in our current analysis of $\hat q$, and will be still important in future EIC.
At lowest order in QCD, the single hadron comes from the fragmentation of the nuclear struck quark, which can rescatter
with the nuclear medium when traversing it. The leading-twist cross section for the single scattering process
can be written as
\bea
\frac{d\sigma^S}{dx_BdQ^2dz_h}\!=&\!\frac{2\pi\alpha^2_{\textrm{em}}}{Q^4}\left[1\!+\!(1-y)^2\right]
\nonumber\\
&\times
\sum_q e_q^2f_{q/A}(x_B,\mu^2) D_{h/q}(z_h, \mu^2)\,.
\label{eq-disxs}
\eea
The $p_T$ broadening in Eq.~(\ref{eq:brd1}), at given values of $x_B$, $Q^2$ and $z_h$, can be expressed as~\cite{Kang:2013raa,Kang:2014ela}
\bea
\hspace{-1.5mm}\Delta \langle p_T^2\rangle \!=\! \left(\!\frac{8\pi^2\alpha_sz_h^2C_F}{N^2_c-1}\!\right)\!
\frac{\sum_q e_q^2T_{qg}(x_B,\!0,\!0,\!\mu^2) D_{h\!/\!q}(z_h,\mu^2)}{\sum_q e_q^2f_{q/A}(x_B,\mu^2) D_{h\!/\!q}(z_h,\mu^2)}\,,
\label{eq-dis}
\eea
where the color factor $C_F$ corresponds to the transport of a quark jet.
In our calculation, the factorization scale in Eq.~(\ref{eq-dis}) is taken to be $\mu^2=Q^2$,
and $\Delta \langle p_T^2\rangle$ is obtained by averaging over a studied kinematic bin.
The theoretical inputs, i.e., the PDFs and FFs, are the same as in our global analysis.

In Fig.~\ref{fig:sidis_xb}, we plot the results for the $p_T$ broadening of pion production in SIDIS in electron-lead~($e$-Pb)
scattering at three EIC facilities, at $Q^2\!=\!2$~GeV$^2$~(left panel) and $10$~GeV$^2$~(right panel), respectively.
The solid curve with shaded band represents the predictions with the kinematic dependent
$\hat q(x,Q^2)$ along with the uncertainties extracted in section~\ref{Hessian}. We find that the $\Delta \langle p_T^2\rangle$ as a function of
$x_B$ clearly reflects the $x$ dependence and the uncertainties of $\hat q(x,Q^2)$, e.g., the growths of both the broadening and the uncertainties
in small- and large-$x$ regions. For comparison, we also show the predictions
with the kinematics independent constant $\hat q$ extracted from the current data as the dashed curve and dotted uncertainty band.
As expected, the differences between the two theoretical predictions appear mainly in small- and large-$x$ regions.
In the bottom of each panel, we show the estimated experimental uncertainties as a reference, which are rather small
compared to the theoretical uncertainties. Therefore, we expect the future EIC experiments should
be able to distinguish the two theoretical predictions and provide powerful constraints on the $\hat q$.

Fig. \ref{fig:sidis_q2} shows the $\Delta \langle p_T^2\rangle$ as a function of the probing scale $Q^2$ in an
intermediate-$x_B$ region where the difference between the extracted $\hat q(x,Q^2)$ and constant $\hat q$ is small.
Since the scale dependence of $\hat q(x,Q^2)$ is mild, we only see small differences between two theoretical
predictions in the studied $Q^2$ regions. The broadening with the constant $\hat q$ slightly
decreases with increasing $Q^2$, which is due to the decreasing averaged $z_h^2$~[a factor in Eq.~(\ref{eq-dis})] as a result of the scale evolution
of the fragmentation function. Similar as in Fig.~\ref{fig:sidis_xb}, the estimated experimental uncertainties
are small for the plotted $Q^2$ region.

\subsection{Nuclear enhancement of di-hadron transverse momentum imbalance}
Now we focus on the nuclear-medium enhanced transverse momentum imbalance of the back-to-back particles production in future EIC.
The transverse momentum imbalance of the particle pair given by $\vec{p}_T=\vec{p}_{1T}+\vec{p}_{2T}$ is equal to
the total transverse momentum of the two particles~\cite{Xing:2012ii}. Thus, the enhancement of the imbalance is also the broadening
of the total transverse momentum.
In this subsection, we study the imbalance of di-hadron~(pion pair) production.
The back-to-back hadron pair comes from the fragmentation of di-jet, which
can be produced through the processes $\gamma^{\ast}q\rightarrow qg$
and $\gamma^{\ast}g\rightarrow q\bar{q}$ at LO in $\alpha_s$.
\begin{figure*}[t]
\includegraphics[width=5.2in]{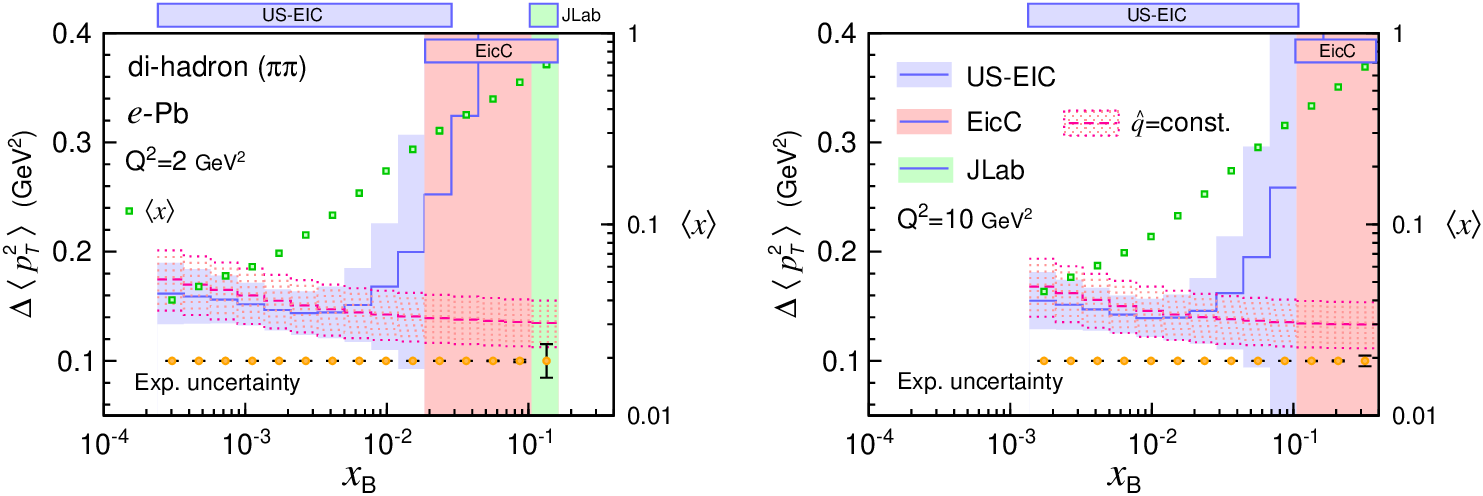}
\caption{Similar as Fig.~\ref{fig:sidis_xb}, but for nuclear enhancement of transverse momentum imbalance $\Delta \langle p_T^2\rangle$ of
di-hadron~($\pi\pi$) production as a function of $x_B$. Result for each bin is calculated in $\gamma^{\ast}$-N system with
fixed $Q^2$ and $x_B$~(central value). Green square marks averaged momentum fraction $x=x_B(1+M_{JJ}^2/Q^2)$
for corresponding $x_B$, with values shown on the right-hand-side vertical axis.
In right panel~($Q^2=10$~GeV), dihadron production in JLab is kinematically prevented/suppressed, thus is not shown.}
\label{fig:dihadron}
\end{figure*}

Following an earlier work~\cite{Xing:2012ii}, the differential cross section of di-hadron production can be written
in the center of mass frame of the virtual-photon-nucleon~($\gamma^{\ast}$-N) system as
\begin{eqnarray}
\frac{d\sigma^S}{dy_1dy_2dp_{1T}dp_{2T}}\!=\!\frac{2\pi\alpha_s\alpha_{\textrm{em}}}{(W^2+Q^2)^2}
\!\sum_{b,c,d}\! D_{h_1/c}(z_1)D_{h_2/d}(z_2)\cr
\times\frac{f_{b/A}(x)}{x} H^U_{\gamma^{\ast}b\rightarrow cd}\,,~~~~~~~~~~~~~~~~~~~~~
\label{eq-dihadron}
\end{eqnarray}
where we have suppressed the scale $\mu^2$-dependence in the PDFs and FFs, $y_{1(2)}$, $p_{1T(2T)}$ and $z_{1(2)}$ are the rapidity, transverse momentum and momentum fraction of
the hadron $h_{1(2)}$, and $H^U_{\gamma^{\ast}b\rightarrow cd}$ represents the perturbatively calculable hard
function of the partonic subprocess $\gamma^{\ast}b\rightarrow cd$~\cite{Xing:2012ii}.
The nuclear enhancement of the transverse momentum imbalance of the di-hadron, induced by the
multiple scattering undergone by the two outgoing partons $c$ and $d$, is only sensitive to
the total color of the two-parton composite state, which is equal to the color of the initial-state
nuclear parton $b$. Accordingly, the nuclear enhancement at given values of $y_{1(2)}$ and $p_{1T(2T)}$ is expressed as~\cite{Xing:2012ii}
\begin{widetext}
\bea
\Delta \langle p_T^2\rangle = \left(\!\frac{8\pi^2\alpha_s}{N^2_c-1}\!\right)
\frac{\sum_{b,c,d} D_{h_1/c}(z_1,\mu^2)D_{h_2/d}(z_2,\mu^2)\frac{1}{x}T_{bg}(x,0,0,\mu^2) H^F_{\gamma^{\ast}b\rightarrow cd}}
{\sum_{b,c,d} D_{h_1/c}(z_1,\mu^2)D_{h_2/d}(z_2,\mu^2)\frac{1}{x}f_{b/A}(x,\mu^2) H^U_{\gamma^{\ast}b\rightarrow cd}}\,,
\eea
\end{widetext}
with the hard function $H^F_{\gamma^{\ast}b\rightarrow cd}$ written as
\bea
H^F_{\gamma^{\ast}b\rightarrow cd}=
\begin{cases}
C_F H^U_{\gamma^{\ast}b\rightarrow cd} ~~~b=\textrm{quark}\\
C_A H^U_{\gamma^{\ast}b\rightarrow cd} ~~~b=\textrm{gluon.}
\end{cases}
\eea
Here the color factor $C_F$~($C_A$) corresponds to the process initiated with a nuclear quark~(gluon). This is more complicated than the case in SIDIS, where only the transport of a quark jet is involved at LO in $\alpha_s$.
In our calculation, the renormalization and factorization scales are taken to be the averaged transverse momentum of di-hadron, i.e., $\mu=(p_{1T}+p_{2T})/2$.
We also employ kinematic cuts as $1<p_{1T(2T)}<4$~GeV and $0.1<y_{1(2)}<2.0$.

In Fig.~\ref{fig:dihadron}, we plot the results for the nuclear enhancement of the di-hadron~(pion pair) imbalance
versus Bjorken $x_B$, at $Q^2\!=\!2$~GeV$^2$~(left panel) and $10$~GeV$^2$~(right panel), respectively.
The predictions with both the kinematics dependent $\hat q(x,Q^2)$ and constant $\hat q$, as well as the estimated
experimental uncertainties are shown.
Note that the momentum fraction $x$ carried by the initial-state nuclear parton $b$ can be expressed as
\bea
x=\frac{Q^2+M_{JJ}^2}{2p_N\cdot\!q_{\gamma}}=x_B\left(1+\frac{M_{JJ}^2}{Q^2}\right),
\label{eq-xdijet}
\eea
where $M_{JJ}$ is the invariant mass of the outgoing two partons~(di-jet) $c$ and $d$. Therefore, the $x$ that enters $\hat q(x,Q^2)$ is larger than the Bjorken variable $x_B$. To show this difference, we mark in Fig.~\ref{fig:dihadron} the averaged momentum fraction $\langle x\rangle$
for the corresponding $x_B$ as green square.
For a same region of Bjorken $x_B$, the di-hadron production probes the $\hat q(x,Q^2)$ at larger values of $x$ compared to the single-hadron production.

\begin{figure*}[t]
\includegraphics[width=5.2in]{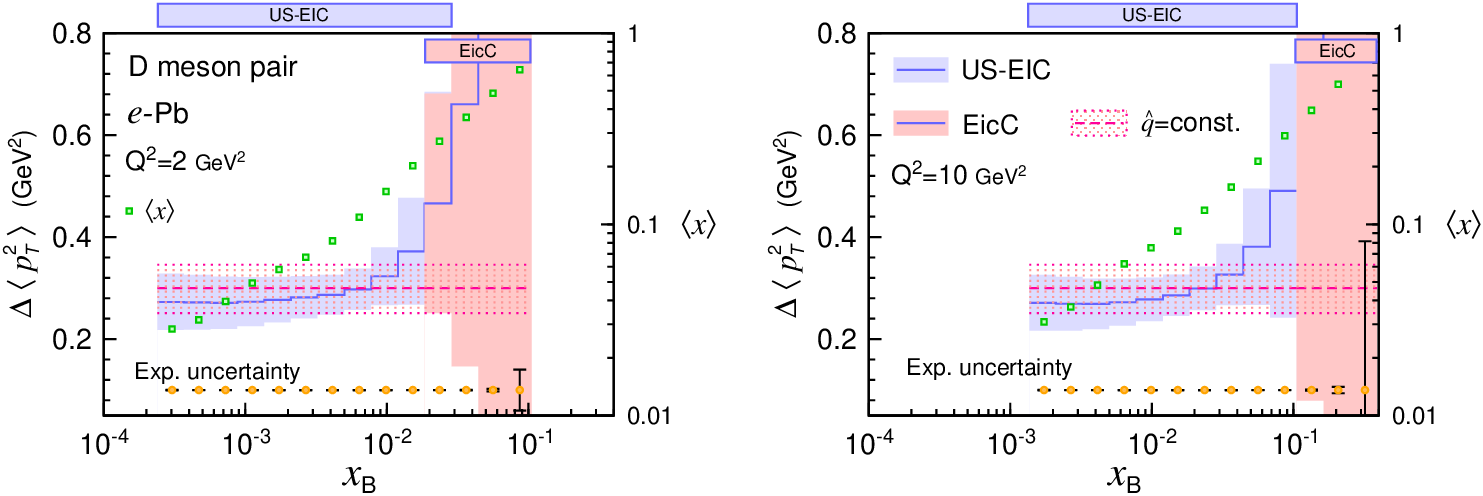}
\caption{Similar as Fig.~\ref{fig:dihadron}, but for nuclear enhancement of transverse momentum imbalance $\Delta \langle p_T^2\rangle$ of
heavy-meson pair~($D\bar{D}$) production as a function of $x_B$. }
\label{fig:DDbar}
\end{figure*}

Since there is a mixture of the processes with color factors $C_A$ ad $C_F$, the di-hadron production is expected to suffer stronger multiple scattering in nuclear medium than the single-hadron production.
At the same time, we can observe the broadening with a constant $\hat q$ become weaker for a greater value of $x_B$~(or $x$) due to the decreasing contributions from the gluon initiated processes.
Since our calculation is made in the $\gamma^{\ast}$-N system, we have
simply estimated the experimental uncertainties with a same luminosity as in $e$-A system to provide a reference.
The measurement of di-hadron imbalance in future EIC is expected to provide
valuable constraints on the behavior of $\hat q(x,Q^2)$ in large-$x$ region, where the large uncertainties from the current analysis make the predictions less reliable.

\subsection{Nuclear enhancement of transverse momentum imbalance of heavy meson pair}
\label{sec-ddbar}

Similar to the di-hadron imbalance, there should be nuclear enhancement of the transverse momentum imbalance
in heavy meson pair production. In this work, we study that for the $D\bar{D}$ pair from the
fragmentation of the produced back-to-back heavy quark pair $Q\bar{Q}$~($c\bar{c}$).
At lowest order in $\alpha_s$, the $c\bar{c}$ di-jet is produced from the photon-gluon scattering
$\gamma^{\ast}g\rightarrow c\bar{c}$.
The differential cross section for $\gamma^{\ast}+A\rightarrow D(p_1)+\bar{D}(p_2)+X$ in the center of mass frame
of the $\gamma^{\ast}$-N system can be expressed in a similar form of the di-hadron production in Eq.~(\ref{eq-dihadron})
as~\cite{Xing:2012ii}
\begin{eqnarray}
\frac{d\sigma^S}{dy_1dy_2dp_{1T}dp_{2T}}=\frac{2\pi\alpha_s\alpha_{\textrm{em}}}{(W^2+Q^2)^2}
D_{D\!/\!Q}(z_1)D_{\bar{D}\!/\!\bar{Q}}(z_2)\cr\cr
\times\frac{f_{g/A}(x)}{x} H^U_{\gamma^{\ast}g\rightarrow Q\bar{Q}}\,.~~~~~~~~~~~~~
\end{eqnarray}
The corresponding nuclear enhancement at given values of $y_{1(2)}$ and $p_{1T(2T)}$ is
\bea
&\Delta \langle p_T^2\rangle = \left(\!\frac{8\pi^2\alpha_s C_A}{N^2_c-1}\!\right)
\frac{T_{gg}(x,0,0,\mu^2)}{f_{g/A}(x,\mu^2) }\,,
\eea
This simple form is related to the fact that the $c\bar{c}$ di-jet
is only initiated with a nuclear gluon, corresponding to the color factor
$C_A$. In our calculation, the fragmentation
from $c$~($\bar c$) quark to $D$~($\bar{D}$) meson is described with KKKS08 fragmentation functions~\cite{Kneesch:2007ey},
and the renormalization and factorization scales are taken to be the averaged transverse mass of the $D\bar{D}$ pair, i.e., $\mu=(m_{1T}+m_{2T})/2$.
We employ the same kinematic cuts as in the calculation for di-hadron production.
The momentum fraction $x$ carried by the initial-state nuclear gluon can be similarly given by Eq.~(\ref{eq-xdijet}).
In heavy meson pair production, the averaged momentum fraction $x$ for a certain value of Bjorken $x_B$ can be even larger than that in di-hadron production, due to the mass of heavy quark.

In Fig.~\ref{fig:DDbar}, we plot the results for the nuclear enhancement of the imbalance of
$D\bar{D}$ pair production versus the Bjorken $x_B$, at $Q^2\!=\!2$~GeV$^2$~(left panel)
and $10$~GeV$^2$~(right panel), respectively. The averaged $\langle x\rangle$ for each kinematic bin is marked in the plot.
It can be observed that, for a same $x$ value, the $\Delta \langle p_T^2\rangle$ in $D\bar{D}$ production is stronger than that in di-hadron production, which is expected as a result of the greater color factor, to be examined in future EIC experiments.

The three types of hard probes in future EIC studied in this section are expected to provide a multi-dimensional
understanding of the jet transport property of cold nuclear matter in a wide kinematic range.
In addition, although the results of $\Delta \langle p_T^2\rangle$ in this section are obtained for the electron-lead
collisions, the results for other colliding nuclei can be easily obtained by rescaling with the nuclear radius~($\times R_A/R_{Pb}$),
according to Eq.~(\ref{eq:Tqg}).

\subsection{Advantage of future EIC measurement for understanding kinematic dependence of $\hat q$}
\label{sec-Reic}
\begin{figure}[b]
\includegraphics[width=3.3in]{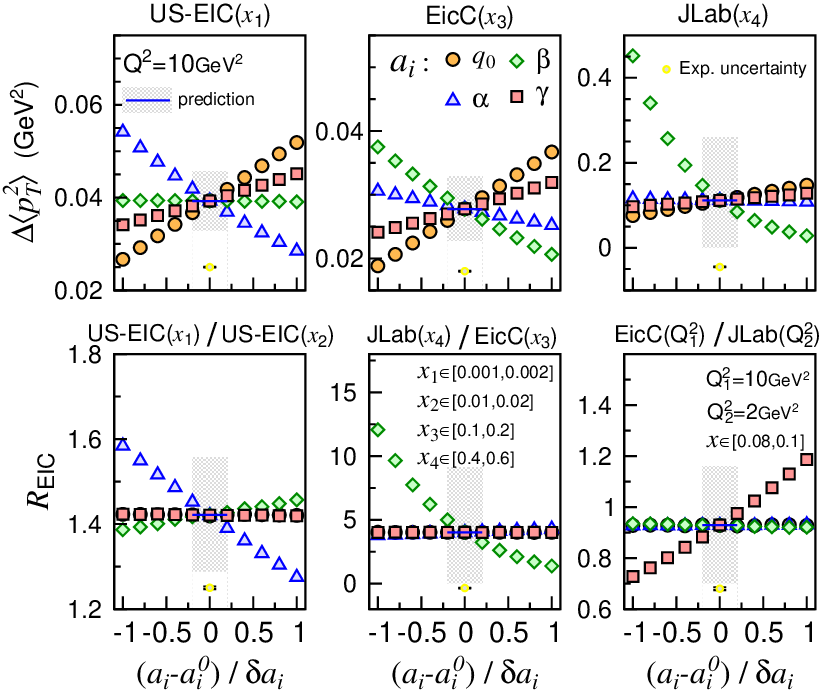}
\caption{Dependence of results on individual parameters, $\hat q_0$, $\alpha$, $\beta$, and $\gamma$, represented with
circle, diamond, triangle, and square symbols, respectively.
Top row: transverse momentum broadening $\Delta \langle p_T^2\rangle$ of single pion production in SIDIS in certain
kinematic regions of three EIC facilities, where $Q^2=$10~GeV$^2$ and $x=x_1$, $x_3$ and $x_4$ for US-EIC, EicC and JLab, respectively.
Bottom row: ratios of $\Delta \langle p_T^2\rangle$ in different kinematic regions defined as Eq.~(\ref{eq-reic}), where $Q^2=$10~GeV$^2$
for left and middle panels, and $x\in[0.08,0.1]$ for right panel.
Blue solid line with grey band represents theoretical prediction with uncertainty.
Abscissa axis shows rescaled parameter values as $(a_i-a_i^0)/\delta a_i$, where $\delta a_i$ is uncertainty of $a_i$.
Yellow circle with vertical bar represents estimated experimental uncertainty.}
\label{fig:eic_r}
\end{figure}

From the above study, we can see that the three future EIC facilities jointly provide a nearly full-kinematics scan of
the jet transport property with high precisions, which will be valuable for determining the kinematic dependence of $\hat q$.
To illustrate this, we can note that, in the sense of the parametrization of $\hat q(x, Q^2)$ in Eq.~(\ref{eq-pramt}),
the kinematic dependence is determined by the parameters $\alpha$, $\beta$, and $\gamma$, and we have the ratio for
two values of $x$
\bea
r(x_1,x_2)\equiv\frac{\hat q(x_1, Q^2)}{\hat q(x_2, Q^2)}=\left(\frac{x_1}{x_2}\right)^{\alpha} \left(\frac{1-x_1}{1-x_2}\right)^{\beta}.
\eea
Assuming that both $\alpha$ and $\beta$ are negative as suggested by our analysis, we can find that,
for small $x$ values~($x_{1,2}\rightarrow0$ and $x_1<x_2$), we have $r(x_1,x_2)\rightarrow(x_1/x_2)^{\alpha}$,
and for large $x$ values~($x_{1,2}\rightarrow1$ and $x_1>x_2$), we have $r(x_1,x_2)\rightarrow[(1-x_1)/(1-x_2)]^{\beta}$.
Thus, the behaviors of $\hat q$ in small and large $x$ regions are dominated by $\alpha$ and $\beta$, respectively.
Similarly, the ratio for two values of $Q^2$
\bea
r(Q^2_1,Q^2_2)\equiv\frac{\hat q(x, Q^2_1)}{\hat q(x, Q^2_2)}=
\frac{\alpha_s(Q^2_1)}{\alpha_s(Q^2_2)}\left[\frac{\ln(Q^2_1/Q^2_0)}{\ln(Q^2_2/Q^2_0)}\right]^{\gamma}
\eea
is only sensitive to the parameter $\gamma$, which dominates the scale dependence.
Accordingly, we can define a secondary observable as a ratio of
$\Delta \langle p_T^2\rangle$ in EIC, such as
\bea
R_{\textrm{EIC}}(x_1,x_2)=\frac{\Delta \langle p_T^2\rangle(x_1, Q^2)}{\Delta \langle p_T^2\rangle(x_2, Q^2)}\,,\cr
R_{\textrm{EIC}}(Q^2_1,Q^2_2)=\frac{\Delta \langle p_T^2\rangle(x, Q^2_1)}{\Delta \langle p_T^2\rangle(x, Q^2_2)}\,,
\label{eq-reic}
\eea
to measure the $x$ and $Q^2$ dependence of $\hat q(x, Q^2)$, respectively.
Figure~\ref{fig:eic_r} shows, for several observables in SIDIS, the dependence of theoretical prediction on each parameters
$a_i$~($\hat q_0$, $\alpha$, $\beta$, and $\gamma$), as a function of the relative parameter displacement
$(a_i-a_i^0)/\delta a_i$, with $\delta a_i$ being the uncertainty of $a_i$.
The three panels in the top row show the dependence for $\Delta \langle p_T^2\rangle$ at three facilities.
We can see a single measurement of $\Delta \langle p_T^2\rangle$ is usually sensitive to more than one parameters.
However, as shown in the bottom row, the three jointly measured $R_{\textrm{EIC}}$ are separately sensitive to the parameters
$\alpha$, $\beta$, and $\gamma$.
The future EIC facilities will allow a precise understanding of the detail of the kinematic dependence of $\hat q$, thanks to the wide
kinematic coverage and high-precision measurement.

\section{Summary and Discussion}
\label{sec-summary}
To gain foreknowledge on how the future electron-ion-collision experiments can deepen our understanding of the jet
transport property of cold nuclear matter, in this work, we study the nuclear-medium induced transverse momentum
broadening/imbalance of single-/pairwise-particle production in electron-ion collisions, for the kinematic
regions of three proposed future facilities.
Our theoretical calculations take into account the multiple scattering undergone by the colored hard
probe that traverses the nucleus, within the framework of the higher-twist expansion, i.e., the
generalized factorization in perturbative QCD.

Particularly, a kinematic dependent jet transport coefficient $\hat q=\hat q(x,Q^2)$, extracted from our global
analysis of the current experimental data, is used in our calculations for EIC.
This globally extracted $\hat q(x,Q^2)$, together with its uncertainty
evaluated with a Hessian matrix method, are available for the community to make theoretical predictions.
Moreover, by adding a new data set on $J/\psi$ production from the LHC, the Hessian analysis results in
reduced uncertainties of $\hat q(x,Q^2)$ in small-$x$ region.
At the same time, we show that this $\hat q(x,Q^2)$ can be converted into the function of jet energy, i.e.,
$\hat q(E_{\textrm{jet}},Q^2)$, which may be instructive for the study of the medium modification
of jet production in heavy-ion collisions.

The current analysis suggests enhancements of $\hat q$ in both small- and large-$x$ regions,
however, the uncertainties in these two regions are still considerable due to the limited data points therein,
which is expected to be better constrained in future EIC experiments.
With the extracted $\hat q(x,Q^2)$, we study three types of observable in EIC, including the transverse
momentum broadening of single hadron production and the enhancement
of two-particle transverse momentum imbalance in di-hadron and heavy-meson pair productions.
These nuclear induced broadening/imbalance are found to be sensitive to the color state of the hard probe,
and to exhibit a clear kinematic dependence stemming from $\hat q(x,Q^2)$. Besides, the results with a constant
$\hat q$ are also given for comparison.
We find that the future EIC experiments have great potential to provide precise understanding of $\hat q$
in a wide kinematic range and to facilitate the jet tomography of cold nuclear medium.

\textit{Note}: The transport coefficient $\hat q(x,Q^2)$ extracted from our global analysis and the uncertainty set are available for user and
can be requested by email from hxing@m.scnu.edu.cn and p.ru@m.scnu.edu.cn.

~

{\it \ Acknowledgments. }\
This research was supported in part
by the National Natural Science Foundation of China (NSFC) under Grants No.~12022512 and No.~12035007, by Guangdong Major Project of Basic and Applied Basic Research No.~2020B0301030008, by Guangdong Basic and Applied Basic Research Foundation~(Project No.~2022A1515110392, 2022A1515010683), 
by the National Science Foundation in US under Grant No.~PHY-1945471~(Z.K.), by the China Postdoctoral Science Foundation under Project No.~2019M652929~(P.R.), and by the MOE Key Laboratory of Quark and Lepton Physics~(CCNU) under Project No.~QLPL201802~(P.R.).

\begin{appendix}
\section{Details of Hessian analysis}
\label{section:appendix}
We briefly review the Hessian analysis employed in this work.
Hessian matrix analysis is a well-known technique for the uncertainty estimation in a global analysis~\cite{Pumplin:2001ct}.
The general idea is to optimize the representation of the parameter space in the neighborhood of the
global minimum $\chi^2$, and to provide an uncertainty set of the parameterized quantity~[e.g. $\hat q(x,Q^2)$ in this work],
from which uncertainties of all related physical quantities can be evaluated.

A Hessian analysis begins with finding in parameter space $\{a_i\}$ the coordinate corresponding to the
minimum of the global $\chi^2$.
The basic assumption of the Hessian analysis is that the $\chi^2$ can be
approximated with a quadratic form in the neighborhood of the minimum~\cite{Pumplin:2000vx,Kovarik:2015cma} as
\bea
\chi^2=\chi_0^2+\sum_{i,j}H_{ij}y_iy_j\,,
\eea
where $\chi_0^2\equiv\chi^2(\{a_i^0\})$ is the global minimum of the $\chi^2$ at the optimal parameter values $\{a_i\}=\{a_i^0\}$,
$y_i=a_i-a_i^0$ is the displacement of $a_i$ from its optimal value $a_i^0$, and $H_{ij}$ is the element of Hessian matrix defined
as the second-order partial derivatives of $\chi^2$ at the minimum
\bea
H_{ij}=\frac{1}{2}\left(\frac{\partial^2\chi^2}{\partial y_i\partial y_j}\right)_{a_i=a_i^0}
\label{eq:Hessian}
\eea
Usually there are interplays among different variables $y_i$~(or $a_i$) in the $\chi^2$, and the off-diagonal Hessian
matrix elements could be non-zero. This makes the uncertainty estimation, corresponding to a certain tolerance
$\Delta\chi^2\equiv\chi^2-\chi^2_0$, not that straightforward.

To disentangle these interplays, one can define a new set of parameters $\{z_i\}$,
in whose representation the Hessian matrix is diagonal. This can be achieved by using the complete set of
$n$ orthonormal eigenvectors $V_i^{(k)}$ of the symmetric $n\times n$ Hessian matrix~($n$ is the number of parameters),
which satisfy the characteristic equations
\bea
\sum_{j} H_{ij}V_j^{(k)}=\lambda_k V_i^{(k)},
\eea
with $\lambda_k$ being the positive eigenvalues. The new parameters $\{z_i\}$ can be expressed as the linear combinations
of the original parameters $\{y_i\}$~\cite{Pumplin:2000vx,Kovarik:2015cma}
\bea
z_i=\sqrt{\lambda_i}\sum_j y_j V_j^{(i)}.
\label{eq-zi}
\eea
With the new set of parameters $\{z_i\}$, the $\Delta\chi^2$ can be written in a simple form as
\bea
\Delta\chi^2=\sum_i z_i^2,
\label{eq-sphere}
\eea
which means that the contours of the $\chi^2$ are spheres in the new basis.
With Eq.~(\ref{eq-sphere}) and the inverse transformation of Eq.~(\ref{eq-zi}), one can
define the uncertainty set $\{a_i^{(\pm k)}\}$ of the original parameters $\{a_i\}$ corresponding to a tolerance $\Delta\chi^2$ as~\cite{Kovarik:2015cma}
\bea
a_i^{(\pm k)}=a_i^0\pm\sqrt{\frac{\Delta\chi^2}{\lambda_k}}V_i^{(k)}\,,~~\textrm{for}~~k=1,2,\ldots,n.
\label{eq-ai}
\eea
For a quantity $\mathcal{Q}$ as a function of $\{a_i\}$, whose value corresponding to $\{a_i^{(\pm k)}\}$ is
$\mathcal{Q}_{\pm k}\equiv\mathcal{Q}(\{a_i^{(\pm k)}\})$, its uncertainty can be evaluated with
\bea
\Delta\mathcal{Q}=\frac{1}{2}\sqrt{\sum^n_{k=1}(\mathcal{Q}_{+k}-\mathcal{Q}_{-k})^2}.
\label{eq-unc}
\eea
As an example, one can give the uncertainty set of $\hat q$
\bea
S_{\pm k}\equiv{\hat q}_{\pm k}={\hat q}(\{a_i^{(\pm k)}\}),
\label{eq-qk}
\eea
and express the uncertainty set for any quantity as a function of $\hat q$ as $\mathcal{Q}_{\pm k}=\mathcal{Q}(S_{\pm k})$.

\begin{figure*}[t]
  \includegraphics[scale=0.6]{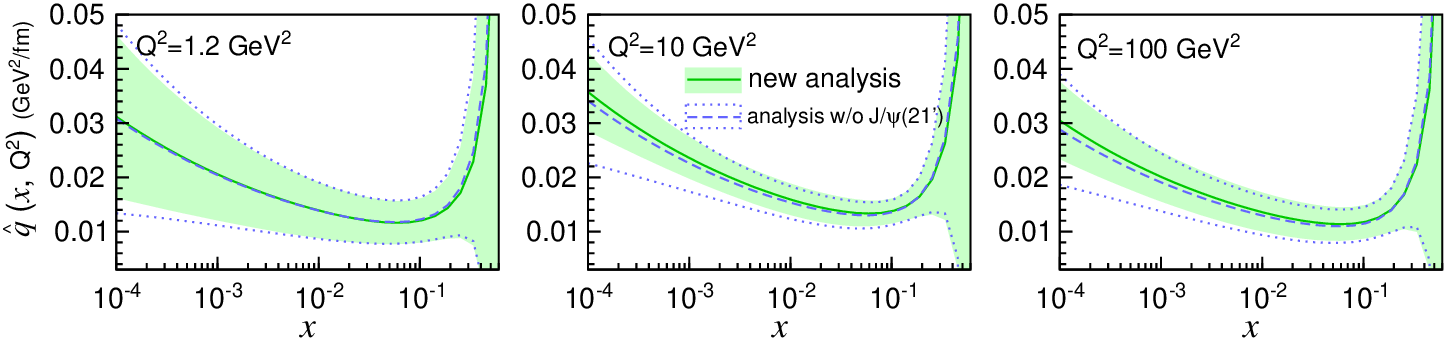}\\
  \caption{$\hat q(x,Q^2)$ and uncertainties extracted from global analysis with and without ALICE data~(2021)
on transverse momentum broadening of $J/\psi$ production. Results of new analysis are the same as in Fig.~\ref{fig:qk_unc}.
Results without $J/\psi$~(21') data are shown with a dashed curve with dotted boundaries.
  }\label{fig:qk_unc_8}
\end{figure*}

In our analysis, the tolerance of the global $\chi^2$ is given at a $p\%$ confidence level~(C.L.) by simply re-scaling the minimum as~\cite{Stump:2001gu,Kovarik:2015cma}
\bea
\Delta\chi^2=\chi^2_0\left(\frac{\xi_{p}}{\xi_{50}}-1\right)\,,
\eea
where the rescaling parameter $\xi_p$ corresponds to the $p$-th percentile of the $\chi^2$ distribution $P(\chi^2,N)$ and is
determined by
\bea
\int_0^{\xi_p}P(\chi^2,N)\,d\chi^2=p\,\%\,,
\eea
with $N$ the number of data points and $P(\chi^2,N)$ defined as
\bea
P(\chi^2,N)=\frac{(\chi^2)^{N/2-1}e^{-\chi^2/2}}{2^{N/2}\Gamma(N/2)}.
\eea
A more thorough and complicated scheme to estimate the tolerance can be found in Refs.~\cite{Stump:2001gu,Kovarik:2015cma}.
In our results, we have estimated the uncertainty at $90$\% C.L., corresponding to the tolerance $\Delta\chi^2=35$.
To get the uncertainty at any other C.L., one can simply rescale the $\Delta\chi^2$ to get the corresponding
uncertainty set $\{a_i\}$ with Eq.~(\ref{eq-ai}).

The Hessian analysis in this work is performed by using the MINUIT package combined with the ITERATE program~\cite{Pumplin:2000vx}, which
will first search for a global minimum of the $\chi^2$ and then calculate the eigenvectors $V_i^{(k)}$ and eigenvalues
$\lambda_k$ of the Hessian matrix with an iterative method~\cite{Pumplin:2000vx}.

\begin{figure*}[t]
  \includegraphics[scale=0.6]{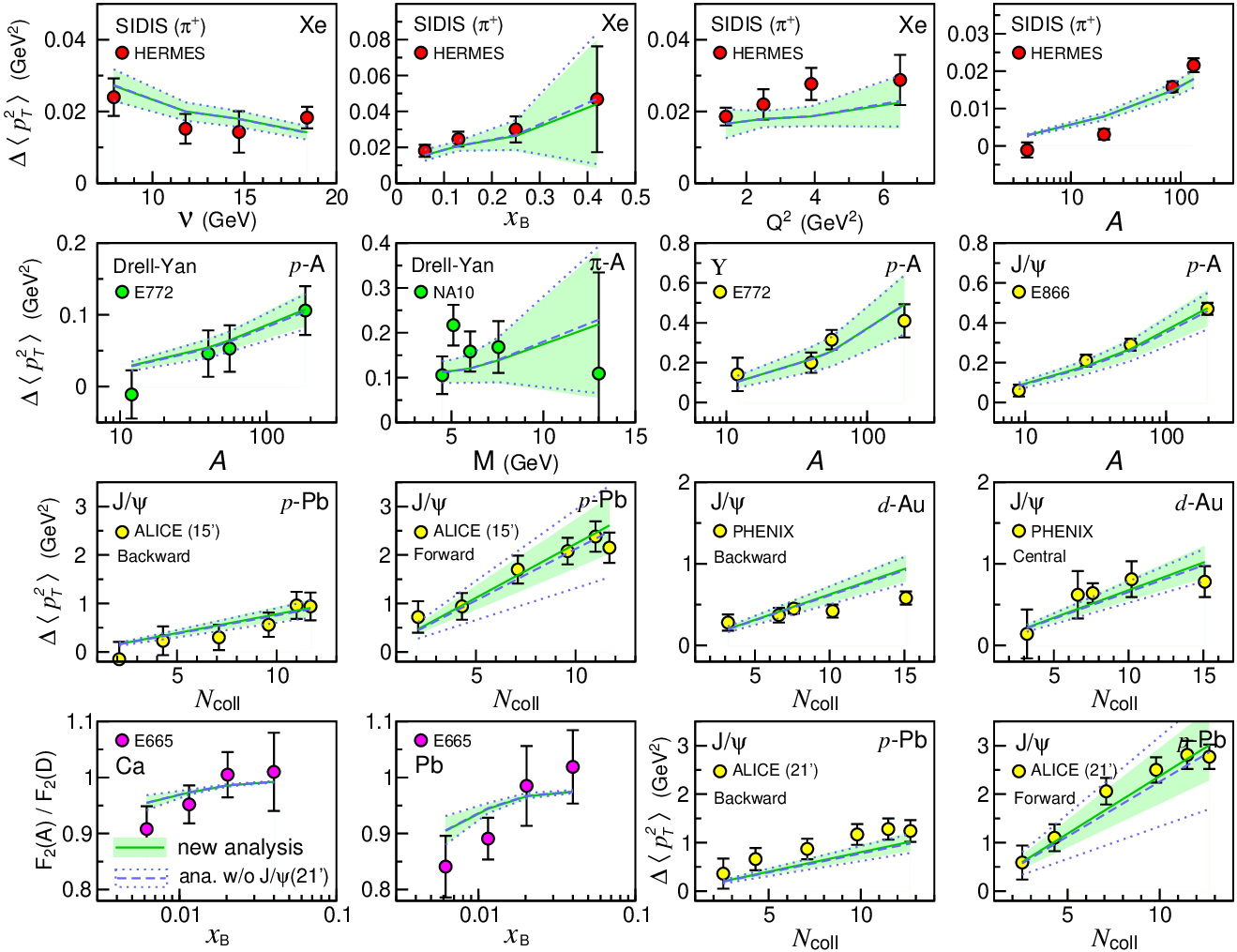}\\
  \caption{Comparison between theoretical results and experimental measurements for part of the representative observables in
  the global analysis of $\hat q$, including transverse momentum broadening $\Delta\langle p_T^2\rangle$ in SIDIS~($e$A),
  Drell-Yan process~($p$A) and heavy quarkonium production~($p$A), and nuclear modification of structure function $F_2(A)/F_2(D)$.
  Theoretical results and uncertainties calculated with $\hat q(x,Q^2)$ extracted in new analysis with ALICE $J/\psi$~(2021) data added are plotted
  with solid curves and shaded areas, while those calculated with $\hat q(x,Q^2)$ extracted by excluding new $J/\psi$~(2021) data are
  plotted with dashed curves and dotted boundaries.
  Circles represent experimental data taken from Refs.~\cite{Airapetian:2009jy,McGaughey:1999mq,Bordalo:1987cr,Alde:1991sw,Peng:1999gx,Leitch:1995yc,Adare:2012qf,Adam:2015jsa,Acharya:2020rvc,Adams:1992nf,Adams:1995is}.
   }\label{fig:Global_Hes}
\end{figure*}

\section{Impact of new ALICE $J/\psi$ data on $\hat q$}
\label{section:appendix2}
In comparison with our previous analysis~\cite{Ru:2019qvz}, we have added a new data set, i.e., the transverse
momentum broadening of $J/\psi$ in $p$-Pb collisions at $\sqrt{s_{NN}}=8.16$~TeV at the LHC~(ALICE 21'), into the
Hessian analysis in this work. To quantitatively show the impact of the new data, we compare the results
with and without including this data set.

In Fig.~\ref{fig:qk_unc_8}, we compare the $\hat q(x,Q^2)$ extracted with and without the new $J/\psi$ data.
We also compare theoretical predictions calculated by using the $\hat q(x,Q^2)$ extracted in the two analyses in Fig.~\ref{fig:Global_Hes}.
In general, the new $J/\psi$ data has small impact on the optimal/central values of the $\hat q(x,Q^2)$ or
theoretical predictions, indicating the new analysis is consistent with our previous work~\cite{Ru:2019qvz}.

With the new $J/\psi$ data, the uncertainties of $\hat q(x,Q^2)$ in small $x$ region are reduced,
especially for $Q^2=10-100$~GeV$^2$~(squared mass of $J/\psi$ $\sim 10$~GeV$^2$), as shown in Fig.~\ref{fig:qk_unc_8}.
As a result, we can see in Fig.~\ref{fig:Global_Hes} that, with the $\hat q(x,Q^2)$ extracted in the new analysis,
the uncertainties of the theoretical predictions for $J/\psi$ production in the
forward region at the LHC are obviously reduced~(2nd panel in 3rd row and last panel in 4th row).
The theoretical uncertainties for other observables, related to mid- and large-$x$ regions, have negligible changes.

\end{appendix}

~

~

~



\end{document}